# Collective relaxation dynamics and vitrification in hexatic smectic binary equimolar mixtures containing component with varying $C_3F_7CH_2OC_mH_{2m}$- chain length

Aleksandra Deptuch[1,*], Mateusz Pączek[2], Anna Drzewicz[1], Marcin Piwowarczyk[1], Magdalena Urbańska[3], Ewa Juszyńska-Gałązka[1,4]

[1] The Henryk Niewodniczański Institute of Nuclear Physics Polish Academy of Sciences, Radzikowskiego 152, PL-31342 Kraków, Poland

[2] Faculty of Chemistry, Jagiellonian University, Gronostajowa 2, PL-30387 Kraków, Poland

[3] Institute of Chemistry, Military University of Technology, Kaliskiego 2, PL-00908 Warsaw, Poland

[4] Research Center for Thermal and Entropic Science, Graduate School of Science, Osaka University, 560-0043 Osaka, Japan

*corresponding author, aleksandra.deptuch@ifj.edu.pl

## Abstract

Differential scanning calorimetry, polarizing optical microscopy, and broadband dielectric spectroscopy are applied for investigation of phase transitions in equimolar binary mixtures consisting of the MHPOBC liquid crystal as one component and 3FmHPhH6 homologs as a second component with different $C_3F_7CH_2OC_mH_{2m}$- chain length (m = 2-7). The paraelectric smectic A* – ferroelectric smectic C* – antiferroelectric smectic $C_A$* – hexatic smectic $X_A$* – glassy smectic $X_A$* sequence is observed for all mixtures. The appearance of the bond-orientational order in smectic $X_A$* is accompanied by an additional relaxation process $P_{hex}$ in dielectric spectra at the low-frequency side of the $P_H$ phason, consistent with its assignment to bond-orientational fluctuations. The empirical exponential-power relationship is proposed for the ratio between the relaxation times of $P_{hex}$ and $P_H$. The behavior of the structural α-relaxation and secondary β-relaxation in smectic $X_A$* depends on the m parameter, changing from overlapping α and β processes for m = 2 to well-separated for m = 6, 7. A double glass transition is observed by calorimetry on heating for m = 5-7. The calorimetric glass transition temperature shows a composition-dependent correspondence with temperatures at which either the $P_H$ or α-relaxation time reaches 100 s, suggesting that different relaxation processes may contribute to vitrification.

## 1. Introduction

Hexatic phases are characterized by the six-fold bond-orientational order (BOO), which combines short-range positional order with long-range orientational order of hexagonally ordered domains. Among liquid crystals, BOO is present in smectic $B_{hex}$, F, and I phases. The molecules are orthogonal to the smectic layers in $SmB_{hex}$, tilted perpendicular to a hexagon edge in SmF, and tilted along a hexagon edge in SmI. The identification of SmF and SmI phases is not always clear, as their very similar structures can be distinguished only for monodomain samples by X-ray diffraction [1]. Thus, the tilted hexatic phases will be further denoted as SmX. In the case of chiral smectic phases, two variants are possible: SmX* and $SmX_A$*, which are hexatic versions of synclinic SmC* and anticlinic $SmC_A$*, respectively [2].

Theoretical model predict collective tilt-magnitude fluctuations and tilt-phase fluctuations in SmC*, which are observed in experimental dielectric spectra as an amplitudon – soft mode (SM) and a phason – Goldstone mode (GM) [2,3]. Description of the SmC*/SmX* transition requires including BOO fluctuations. The theoretical model presented in Ref. [4] assumes that they have lower frequencies than corresponding tilt fluctuations both in SmC* and SmX*. In SmC*, where BOO is short-range, related phason and amplitudon have similar frequencies. Experimentally, the Goldstone mode provides the main contribution to the dielectric response in SmC* [5-7], although the BOO fluctuations can also be detected, particularly on approaching the SmC*/SmX* transition [4]. In SmX*, BOO fluctuations are predicted to have a larger contribution to the dielectric response than in SmC*, with the phason stronger than the amplitudon. There is also a shift of the phason and amplitudon to a lower and higher frequency, respectively. The increase in dielectric strength and decrease in frequency of the BOO phason at the SmC*/SmX* transition were confirmed experimentally for some compounds [4,8]. Similarly, Ref. [9] shows a pronounced shift toward lower frequencies of a process in the SmX* phase on cooling.

The dielectric spectra of $SmC_A$* are usually recognizable by two processes, low-frequency $P_L$ and high-frequency $P_H$ [10] (although they are absent for some compounds [11]). They can be interpreted as tilt phasons, where fluctuations occur in-phase ($P_L$) and anti-phase ($P_H$) in neighboring smectic layers [10]. As $P_L$ shows the Arrhenius behavior, it is sometimes attributed to the rotation around the short molecular axis (s-process) [12-14]. Results in Refs. [10,15] indicate that both in-phase fluctuations and s-process may contribute to $P_L$. The $SmC_A$*/$SmX_A$* transition is often characterized by the appearance of a new process $P_{hex}$ at a frequency between $P_L$ and $P_H$ [16-21]. The new process can be interpreted as the BOO phason [11], although in Ref. [21], $P_{hex}$ was considered as the s-process. In the MHPOBC and MHPB(H)PBC compounds, the $SmC_A$* → $SmX_A$* transition results in the slowing down of the $P_L$ and $P_H$ processes without the appearance of any additional process between them [15,22]. In a mixture from Ref. [23], $P_H$ slows down and a new process, interpreted as the s-process, appears at higher frequencies. Meanwhile, in the MHPNBC and MHPB(F)PBC compounds, the $P_H$ relaxation time decreases in the $SmX_A$* phase [13,14]. Reports also differ regarding the relaxation processes associated with the glass transition of $SmX_A$*. Compounds from Refs. [16-18] show separated primary α-relaxation and secondary β-relaxation, while for a mixture from Ref. [19], the α- and β-relaxations seem to overlap. Thus, the available results indicate the need for a broader study of relaxation processes in $SmX_A$* and its glass.

This paper aims to investigate six equimolar binary mixtures of the MHPOBC compound and 3FmHPhH6 homologs, m = 2-7 (Figure 1), with a focus on the dielectric relaxation processes in $SmC_A$* and $SmX_A$*, as well as the glass transition of $SmX_A$* in connection with the $C_mH_{2m}$ chain length of the fluorinated 3FmHPhH6 component. The systematic variation of a single molecular structural parameter provides an opportunity to examine how the terminal chain length affects collective relaxation dynamics, the separation of structural and secondary relaxation processes, and the vitrification behavior of the hexatic phase.

MHPOBC was the first liquid crystal with proven antiferroelectric properties [24]. The isotropization and melting temperatures of MHPOBC are 421 K and 357 K, respectively [25]. MHPOBC forms SmA* and $SmC_A$* phases. The phase sequence in a narrow temperature range between SmA* and $SmC_A$* depends on the enantiomeric purity; $SmC_\alpha$*, SmC*, $SmC_{FI1}$*, or $SmC_{FI2}$* phases can be observed [25,26]. Supercooled MHPOBC forms the hexatic $SmX_A$* phase below 338 K [25]. 3FmHPhH6 homologs form SmA* (m = 6, 7), SmC* (m = 4, 5, 6, 7), and $SmC_A$* (all homologs) phases. The isotropization and melting temperatures of 3FmHPhH6 are 389-403 K and 331-352 K, respectively [27]. The transition to $SmX_A$* upon supercooling occurs for m = 2, 3, 4, 5, 7 at 263-301 K [16,17,19,28]. 3FmHPhH6 homologs with m = 5, 7 form the glass of $SmX_A$* below 247 K and 230 K instead of a crystal phase at moderate cooling rates [16,17].

Based on these findings, the hexatic $SmX_A$* phase can be expected in the mixtures of MHPOBC and 3FmHPhH6, possibly undergoing the glass transition. In the first part, the phase sequence in mixtures is determined by differential scanning calorimetry (DSC) and polarizing optical microscopy (POM). In the second part, dielectric relaxation processes and ionic conductivity are studied by broadband dielectric spectroscopy (BDS). The glass transition temperatures determined by DSC and BDS methods are discussed. In the third part, the relationship between relaxation times of $P_{hex}$ and $P_H$ in $SmX_A$* is analyzed and compared to literature reports.

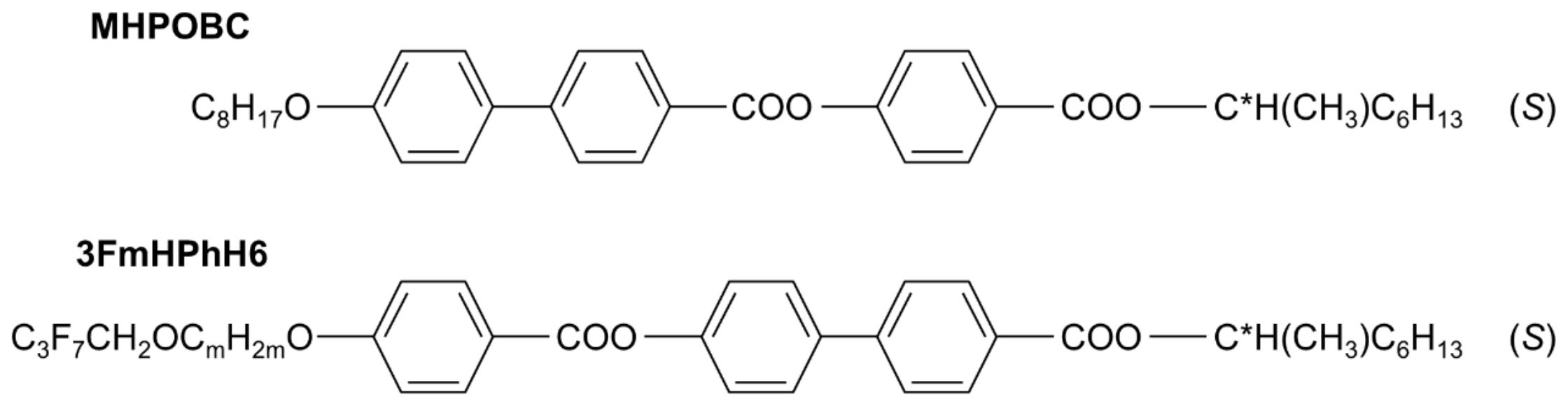

Figure 1. Molecular structures of the MHPOBC compound and 3FmHPhH6 homologs (m = 2-7).

## 2. Experimental details

The MHPOBC compound [29] and 3FmHPhH6 homologs (m = 2, 3, 4, 5, 6, 7) [27], all S-enantiomers, were synthesized in the Institute of Chemistry of the Military University of Technology in Warsaw. Six binary equimolar MHPOBC: 3FmHPhH6 mixtures, denoted as MIXmHH6, were prepared by dissolution of components in acetone followed by solvent evaporation and heating of the resulting material up to 433 K, to the isotropic liquid phase, in order to ensure total evaporation of acetone and good mixing of components.

The DSC measurements were carried out with the DSC 2500 TA Instruments calorimeter. The samples weighting 7.09, 6.44, 9.06, 7.76, 7.14, 7.01 mg for MIXmHH6 with m = 2, 3, 4, 5, 6, 7, respectively, were sealed in the aluminum pans. Each sample was initially heated from the room temperature to isotropic liquid at 10 K/min (1st heating). Then, measurements were performed during cooling and subsequent heating with rates 2, 5, 10, 15, 20, 25, 30 K/min in the 173-443 K range. The DSC thermograms were analyzed with TRIOS and OriginPro.

The POM textures were collected on cooling and heating at 10 K/min in the 188-433 K range with the Leica DM2700 P polarizing microscope. The samples were placed between two thin glass slides without aligning layers. The mean weighted intensity and the mean red, green, blue (rgb) components of each image were calculated in TOApy [30] and ImageJ [31].

The BDS spectra were collected with the Novocontrol impedance spectrometer for samples between gold electrodes with polytetrafluoroethylene spacers. The sample thickness was ca. 50 μm and the electrode's active area had a diameter of 10 mm. The BDS measurements were performed on cooling and heating between 423 K and 173 K over a frequency range of 0.1 Hz – 10 MHz, using a temperature step of 2 K or 1 K. The spectra were analyzed in OriginPro.

## 3. Results and discussion

### *3.1. Phase sequence and glass transition*

The DSC thermograms, along with the corresponding phase transition temperatures obtained at different cooling/heating rates, are presented in Figures S1 and S2 of the supplementary materials. The melting endotherms of a crystal phase on the 1st heating are shown in Figure S3. The DSC results, summarized in Figure 2 and Table S1, show the same phase sequence on cooling for all mixtures: Iso → SmA* → SmC* → $SmC_A$* → $SmX_A$* → glassy $SmX_A$* (gl$SmX_A$*). The odd-even effect occurs, with the temperature range of SmC* broader for mixtures with even m. The SmA*, SmC*, $SmC_A$* phases are enantiotropic, while $SmX_A$* is monotropic for all mixtures, as the $SmC_A$*/$SmX_A$* transition temperature is lower than the melting temperature. The cooling rate 2 K/min is enough to obtain gl$SmX_A$* instead of a crystal phase. The glass transition temperature $T_g$ depends on the length of the $C_mH_{2m}$ chain. It initially increases from 264 K for m = 2 to 270 K for m = 3, then decreases with increasing m down to 231 K for m = 7. The behavior on heating also depends on the $C_mH_{2m}$ chain length. For m = 2, the cold crystallization occurs for all heating rates, with the onset at 282-297 K, and the crystal phase melts at 316 K. For m = 3, the exotherm from the cold crystallization is visible at 293 K for 2 K/min, but the melting endotherm is observed at 315-319 K up to 10 K/min. For m = 4, the exotherm from the cold crystallization is not visible, possibly overlapping with the endotherm from the $SmX_A$*/$SmC_A$* transition, but the melting endotherm is present at 315 K for 2 and 5 K/min. No signs of the cold crystallization and melting are observed for m = 5, 6, 7 at any heating rate. Instead, distinct calorimetric glass-transition steps are detected for these samples on heating, despite a single resolved glass-transition step on cooling. This behavior may indicate the dynamic arrest of different molecular degrees of freedom, although thermal history effects or dynamic heterogeneity cannot be excluded. The first glass transition occurs at $T_{g1}$ = 224-226 K and the second one at $T_{g2}$ = 254, 250, 246 K for m = 5, 6, 7, respectively. For m = 5, $T_g$ obtained on cooling is only slightly lower than $T_{g2}$, while for m = 6, 7, $T_g$ from cooling lies between $T_{g1}$, $T_{g2}$ from heating.

The fan-shaped textures of smectic phases observed by POM are presented in Figures S4-S15. Transitions between SmA*, SmC*, and $SmC_A$* are visible as abrupt color changes. Meanwhile, the $SmC_A$*/$SmX_A$* transition leads to a smoother change, as it is related only to an increasing range of BOO instead of appearance (SmA*/SmC*) or alteration (SmC*/$SmC_A$*) of the tilt order. The glass transition of $SmX_A$* is not visible in POM textures, indicating that vitrification does not produce detectable changes in the optical textures under the experimental conditions.

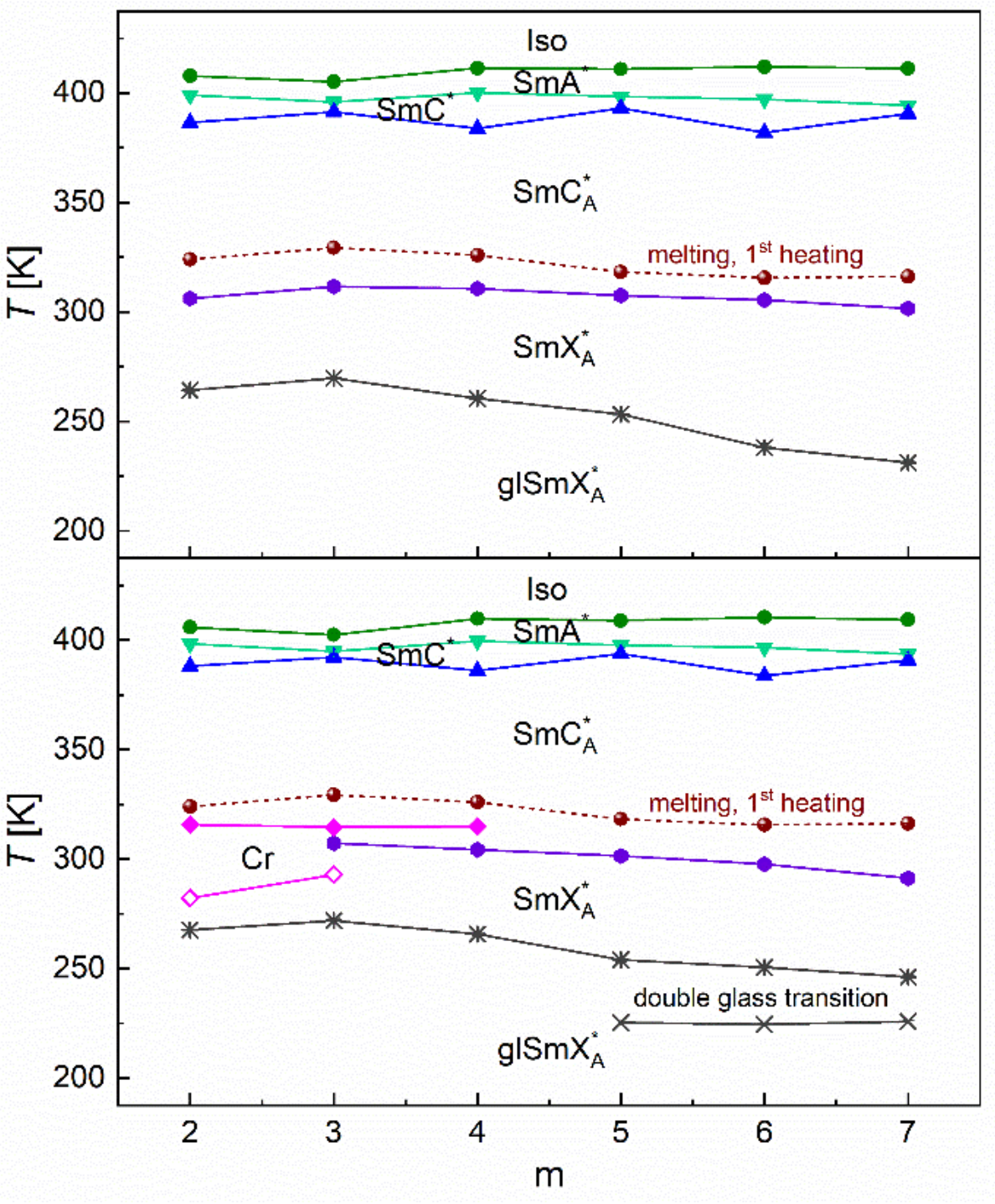

Figure 2. Phase diagram of MIXmHH6 as a function of the $C_mH_{2m}$ chain length obtained from DSC thermograms. The upper and bottom panels show the results from cooling and heating, respectively. Melting temperature from the 1st heating indicates a boundary between stable and supercooled $SmC_A$*.

### 3.2. Arrhenius and super-Arrhenius behavior of dielectric relaxation processes

The BDS spectra were analyzed by the Havriliak-Negami model [2,32,33] (Equation (1), Figure S16), where: $\varepsilon^*$ is a complex dielectric permittivity with a real $\varepsilon'$ and imaginary $\varepsilon''$ parts, corresponding to dielectric dispersion and absorption, $\varepsilon_\infty$ is dielectric dispersion at high frequencies, $\Delta\varepsilon$ is dielectric strength, $\tau_{HN}$ is Havriliak-Negami relaxation time, $a$ and $b$ are shape parameters, $S_1, S_2, n_1, n_2$ describe low-frequency background:

$$\varepsilon^*(f) = \varepsilon'(f) - i\varepsilon''(f) = \varepsilon_\infty + \sum_j \frac{\Delta\varepsilon_j}{\left(1+\left(2\pi i f \tau_{HNj}\right)^{1-a_j}\right)^{b_j}} - \frac{iS_1}{(2\pi f)^{n_1}} + \frac{S_2}{(2\pi f)^{n_2}}. \quad (1)$$

The relaxation time $\tau$ corresponding to the peak position in $\varepsilon''$ was calculated from Equation (2) if $b \neq 1$ [32]:

$$\tau = \tau_{HN}\left(\sin\left(\frac{\pi(1-a)}{2+2b}\right)\right)^{-\frac{1}{1-a}}\left(\sin\left(\frac{\pi(1-a)b}{2+2b}\right)\right)^{\frac{1}{1-a}}. \quad (2)$$

The activation plots of $\tau$ are presented in Figure 3. Most of relaxation processes are observed in all mixtures. The soft mode (SM), recognized by $\tau$ increasing on cooling [3], is detected in the SmA* phase (Figure S16a). The strong Goldstone mode (GM) [3] characterizes the SmC* phase (Figure S16b). The low-frequency in-phase phason ($P_L$) and high-frequency anti-phase phason ($P_H$) [10] characterize the SmC$_A$* phase (Figure S16c). The $P_L$ and $P_H$ processes are still present in the SmX$_A$* phase, with an additional $P_{hex}$ process of intermediate frequency, splitting from $P_H$ at the SmC$_A$*/SmX$_A$* transition. $P_{hex}$ is tentatively assigned to the BOO phason, consistently with previous dielectric studies [11] (Figure S16d). The low-frequency Maxwell-Wagner-Sillars (MWS) process, associated with interfacial charge accumulation [34], is observed in SmA*, SmC*, and SmC$_A$*. The α-relaxation, related to the glass transition [32], is present on the high-frequency side of $P_H$ in SmX$_A$* (Figure S16e), followed by secondary β-relaxation [35,36] and γ-relaxation [36,37] in the glSmX$_A$* state (Figure S16f).

An additional $P_{hex}$’ process is observed for MIX2HH6 on cooling (Figure S16g). $P_{hex}$’ may represent a shifted $P_{hex}$ relaxation associated with SmX$_A$* regions affected by partial crystallization during cooling. It is indicated by the overlap of the relaxation times of $P_{hex}$’ with that of the $P_{hex}$ process on heating, where the fraction of the crystal phase is larger. There is another X process between $P_H$ and α in the SmX$_A$* phase of MIX3HH6 (Figure S16h). The origin of the X-process is not certain, it may arise from the crystallized fraction of the sample.

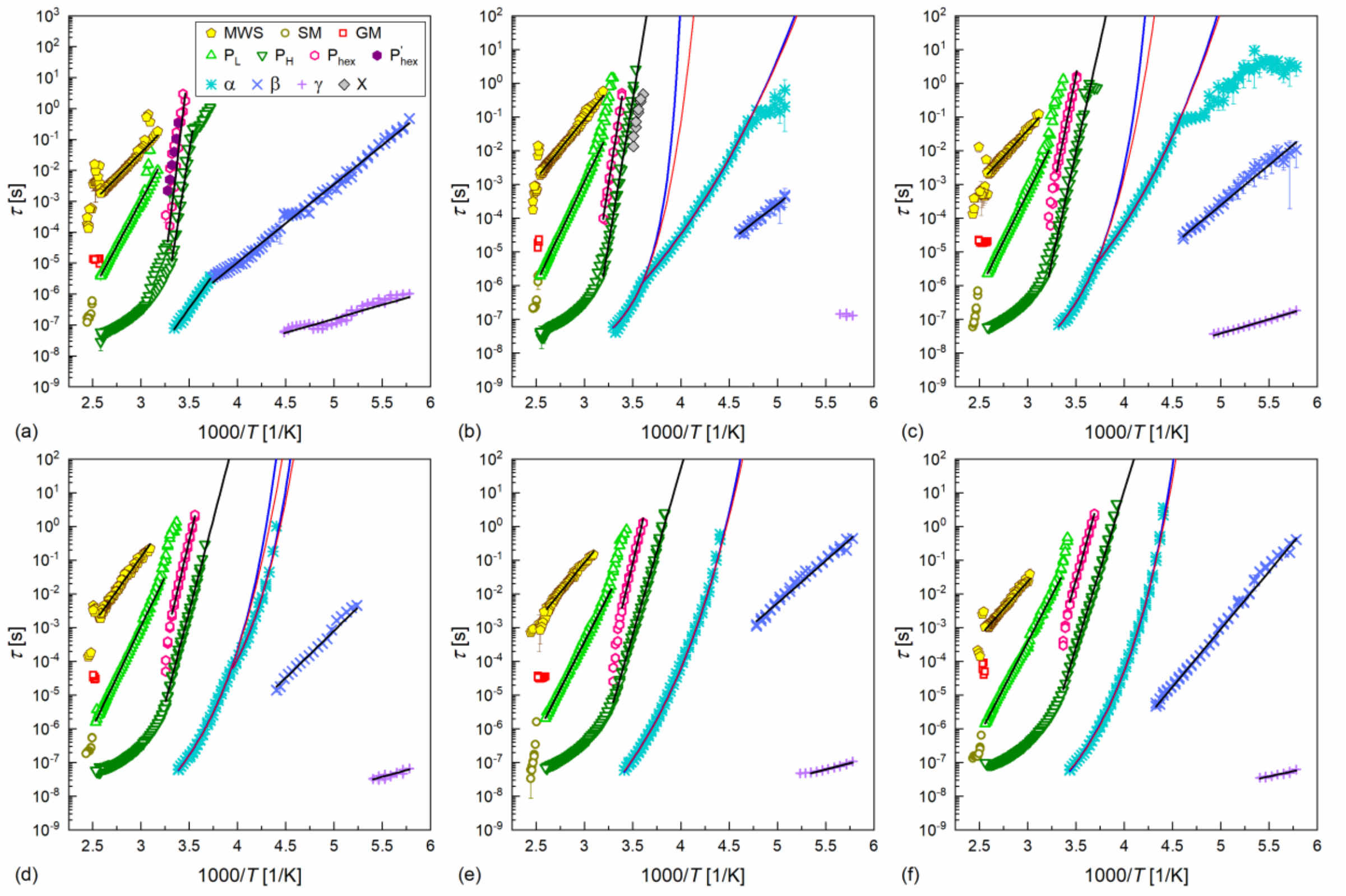

Figure 3. Activation plots of relaxation processes in MIXmHH6 with m = 2 (a), 3 (b), 4 (c), 5 (d), 6 (e), 7 (f). Points – experimental values; black, blue, and red lines – fitting results of the Arrhenius, VFT, and MYEGA formulas, respectively. The legend in (a) is common for all panels.

The Arrhenius relation of the relaxation time, $\tau(T) = \tau_0 \exp{(E_a/RT)}$, can be applied to MWS, $P_L$ in $SmC_A$*, $P_{hex}$, $P_H$ in $SmX_A$*, and β, γ processes in gl$SmX_A$*. The ionic conductivity, determined as $\sigma_{ion} = S_1/\varepsilon_0$ if $n_1$ = 1 [32], also shows the Arrhenius dependence in $SmC_A$* (Figure 4). The activation energies are summarized in Figure 5 and Table S2. The $E_a$ of MWS process and $\sigma_{ion}$ are equal to 60-77 kJ/mol and 54-71 kJ/mol, respectively, and they are strongly correlated, with the 1.109(7) scaling factor obtained from the linear fit (Figure S17). The theoretical description from Ref. [34] indicates proportionality between $\sigma_{ion}$ and MWS relaxation time, which is consistent with the observed correlation and supports the assignment of the low-frequency process to charge-accumulation effects. The $P_L$ process has $E_a$ of 105-115 kJ/mol, without any monotonic dependence on m. $E_a$ ~ 100 kJ/mol is often reported for liquid crystals with similar molecular structures [12,20,13,15,38] and agrees with $E_a$ of the s-process determined for a homeotropically aligned sample in Ref. [15]. This supports a contribution of the s-process to $P_L$, next to the in-phase tilt fluctuations. The largest $E_a$ values are 197-504 kJ/mol for $P_{hex}$ and 178-379 kJ/mol for $P_H$; in both cases $E_a$ decreases with increasing m. For m = 4-7, the Arrhenius dependence can be fitted to $\tau_{P_{hex}}$ only deep into the $SmX_A$* phase; the exact $\tau_{P_{hex}}$ and $\tau_{P_H}$ relationship will be presented in the next section. The β process has $E_a$ in the 43-66 kJ/mol range. It can originate either for movements of a whole molecule (Johari-Goldstein process) or only its part [35,36]. The latter may be plausible for the flexible MHPOBC and 3FmHPhH6 molecules, although a Johari-Goldstein contribution cannot be excluded on the basis of the present data. The energy barriers calculated by DFT method in Refs. [16,18] indicate that the process with $E_a$ of ca. 50 kJ/mol may originate, e.g., from rotations of the aromatic ring. The γ process has the lowest $E_a$ of 12-17 kJ/mol; for m = 3 there were too few experimental points to determine $E_a$ but it is supposed to be low, as the slope in the activation plot is close to zero (Figure 3b). The γ process may be associated with localized motions of terminal alkyl-chain segments [39], although its specific molecular origin remains to be established.

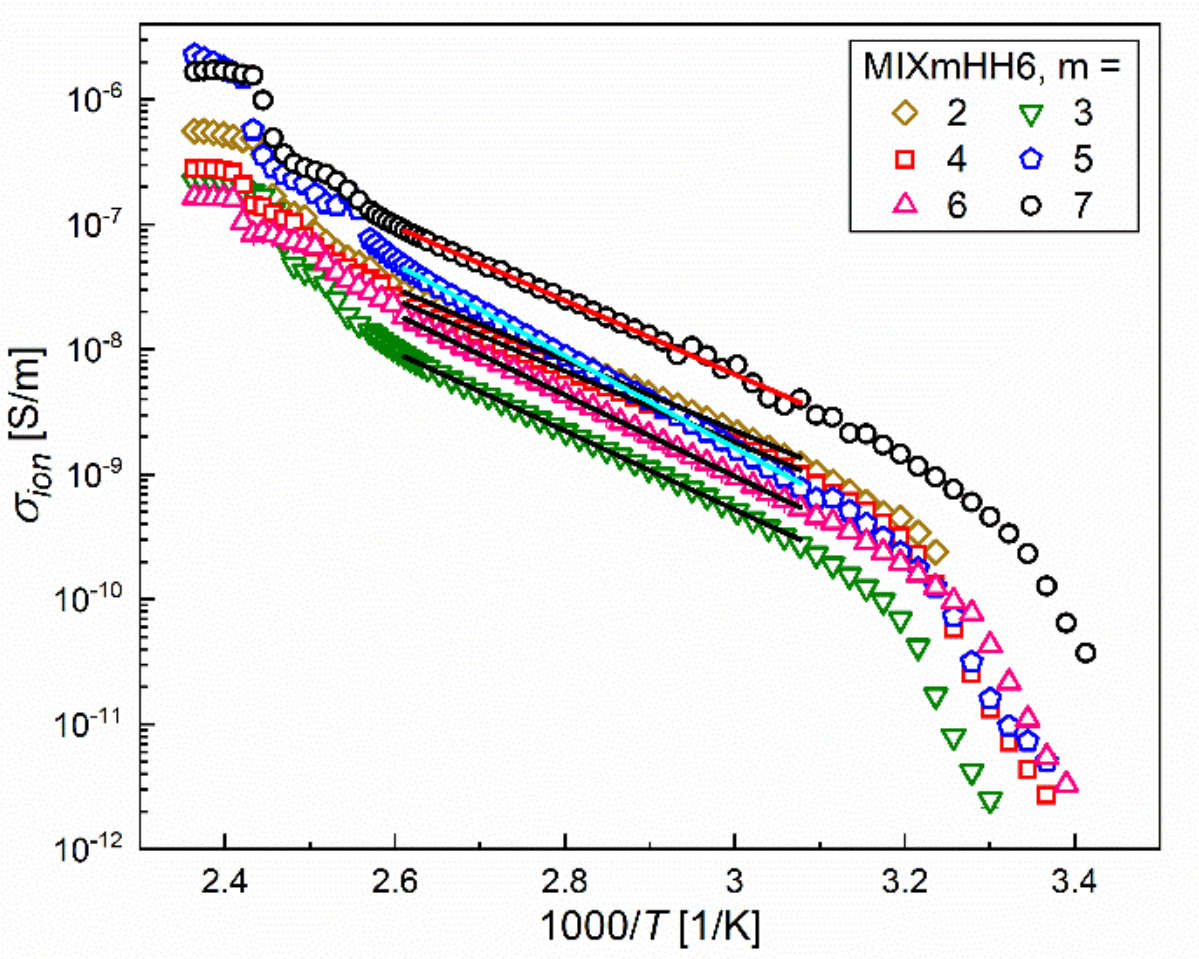

Figure 4. Activation plot of ionic conductivity in MIXmHH6 on cooling.

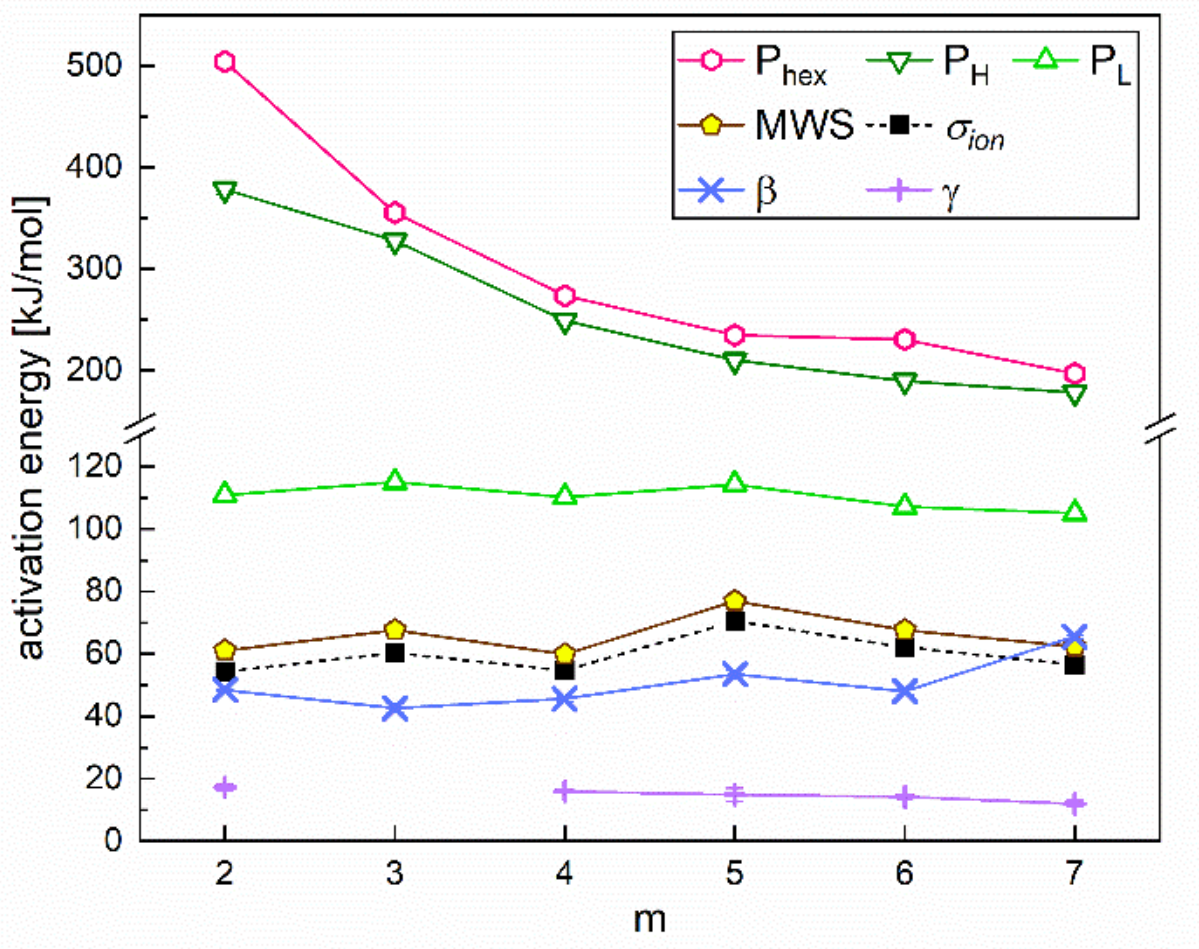

Figure 5. Activation energies of relaxation processes and ionic conductivity in MIXmHH6.

The behavior of the α-relaxation depends strongly on m. MIXmHH6 with m = 6, 7 show the super-Arrhenius behavior of the α-relaxation time (Figure 3e,f), which can be described either by Vogel-Fulcher-Tammann (VFT, Equation (3)) [40] or Mauro-Yue-Ellison-Gupta-Allan (MYEGA, Equation (4)) [41] formulas:

$$\tau_\alpha(T) = \tau_\infty \exp\left(\frac{B}{T-T_V}\right), \tag{3}$$

$$\tau_\alpha(T) = \tau_\infty \exp\left(\frac{K}{T}\exp\left(\frac{C}{T}\right)\right), \tag{4}$$

where: $\tau_\infty$ is pre-exponential constant, $B$, $K$, $C$ are parameters with a unit of temperature, proportional to energy barriers, $T_V$ is the Vogel temperature where the α-relaxation time diverges. The deviation from the Arrhenius dependence is described by the fragility index $m_f$, defined by Equation (5) [40]:

$$m_f = \left.\frac{\mathrm{d}\log_{10}\tau_\alpha(T)}{\mathrm{d}(T_g/T)}\right|_{T=T_g}, \tag{5}$$

where $\tau_\alpha(T_g)$ = 100 s. The α-relaxation time for m = 3, 4, 5 is described by two sets of VFT and MYEGA parameters for low and high temperatures (Figure 3b-d). Upon appearance of the β-relaxation, the apparent α-relaxation time for m = 3 and 4 deviates from the super-Arrhenius dependence and approaches an Arrhenius-like temperature dependence with an activation energy comparable to that of the β-relaxation. However, the data are scattered due to the overlap between α and β processes. Thus, $\tau_\alpha$ values in there ranges may be biased and they are not analyzed. Finally, $\tau_\alpha$ has an Arrhenius dependence for m = 2, where the α-relaxation is observed in a narrower temperature range than for larger m (Figure 3a).

The $m_f$ and $T_g$ values of MIXmHH6 are presented in Figures 6 and 7, respectively, as well as in Table S3. The fragility index obtained by VFT is significantly higher and has a larger uncertainty than obtained by MYEGA in the high-temperature region for m = 3, 4; differences are caused by a fitting in a rather narrow temperature range, well above $T_g$. In other cases, VFT and MYEGA results are closer to each other, with higher $m_f$ determined by VFT. The qualitative dependence of $m_f$ on the $C_mH_{2m}$ chain length is the same for both models. For m = 3, the fragility in the high-temperature region is much larger than in the low-temperature region. As m increases to 5, both $m_f$ values approach each other, with a decrease at high $T$ and increase at low $T$. Finally, they are equal for m = 6, 7, where a single super-Arrhenius dependence is observed. The $m_f$ of m = 2 was attributed to the low-$T$ region based on $T_g$ and it agrees with the $m_f$ trend for higher m. The deviation from the Arrhenius dependence is moderate in the low-$T$ region, as the fragility is below 100 for all MIXmHH6. Very high apparent fragility with $m_f$ > 100 is obtained in the high-$T$ region for m = 3, 4; the latter for the VFT equation only. These values should be interpreted cautiously because they are derived from fits over a limited temperature interval and may be sensitive to the chosen relaxation model. By extrapolation of $m_f$, $T_g$ values for m = 3, 4, it is supposed that the high-$T$ region with a large fragility may exist also for m = 2, although its temperature range is too narrow to be distinguished from the experimental activation plot.

The dynamic glass transition temperature for m = 2-4 at $\tau_\alpha$ = 100 s is lower than $T_g$ from DSC results (Figure 7). At the same time, good agreement with calorimetric $T_g$ is obtained for extrapolation of the $P_H$ relaxation time to 100 s. For m = 5-7, the lower and higher calorimetric $T_g$ of a two-step glass transition on heating correspond to $\tau_\alpha$ = 100 s and $\tau_{P_H}$ = 100 s, respectively. On cooling, a one-step glass transition in DSC results is observed for m = 5-7, with $T_g$ corresponding to $\tau_{P_H}$ = 100 s for m = 5 and close to $\tau_\alpha$ = 100 s for m = 7, with an intermediate value for m = 6. These results suggest a composition-dependent relationship between the calorimetric glass transition and the slowing down of $P_H$ and α-relaxation processes. For smaller m, the calorimetric $T_g$ is closer to the temperature at which the $P_H$ relaxation time reaches 100 s, whereas for larger m, an additional correspondence with the α-relaxation is observed. However, the relative contributions of these processes to the heat-capacity step cannot be quantified from the present data.

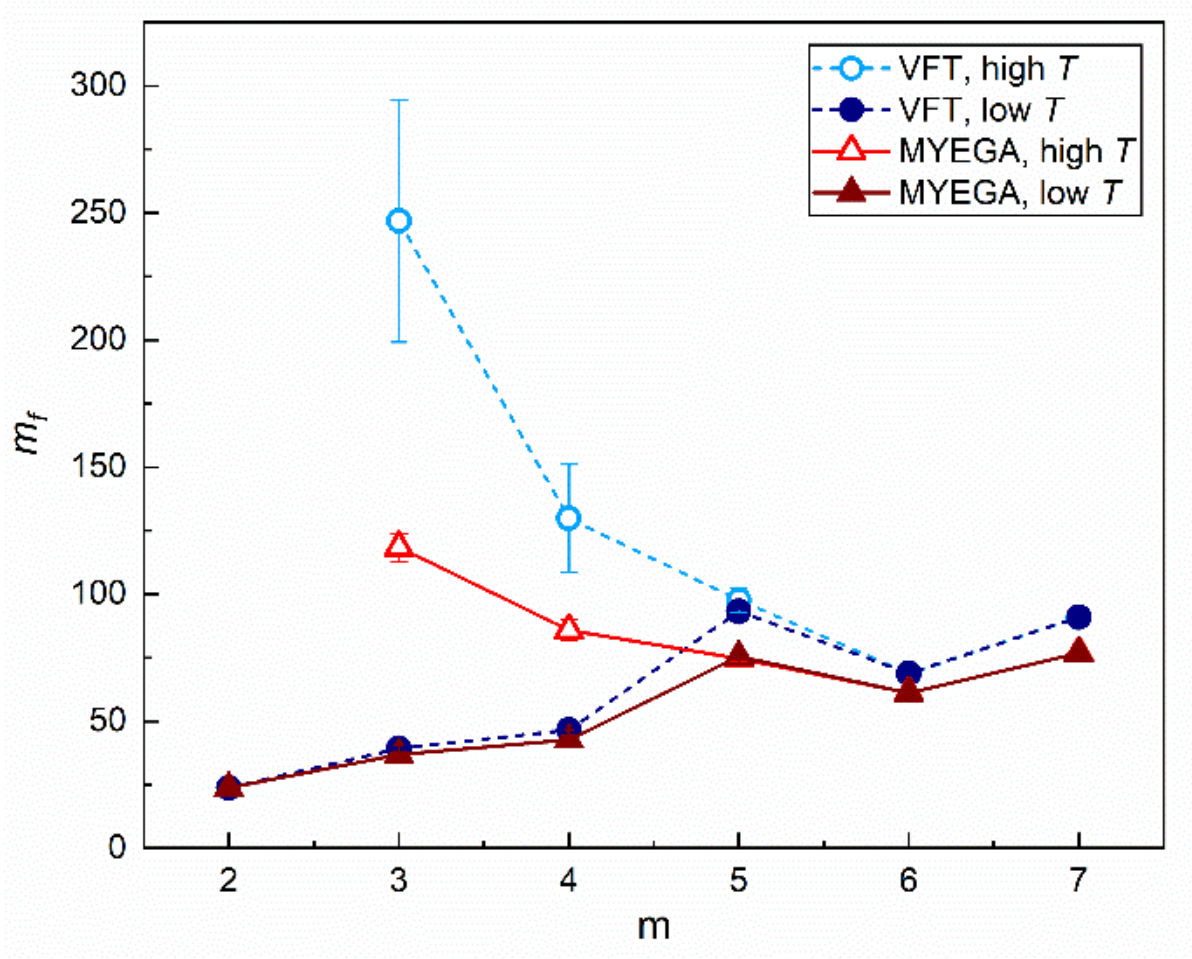

Figure 6. Fragility index of α-relaxation in MIXmHH6.

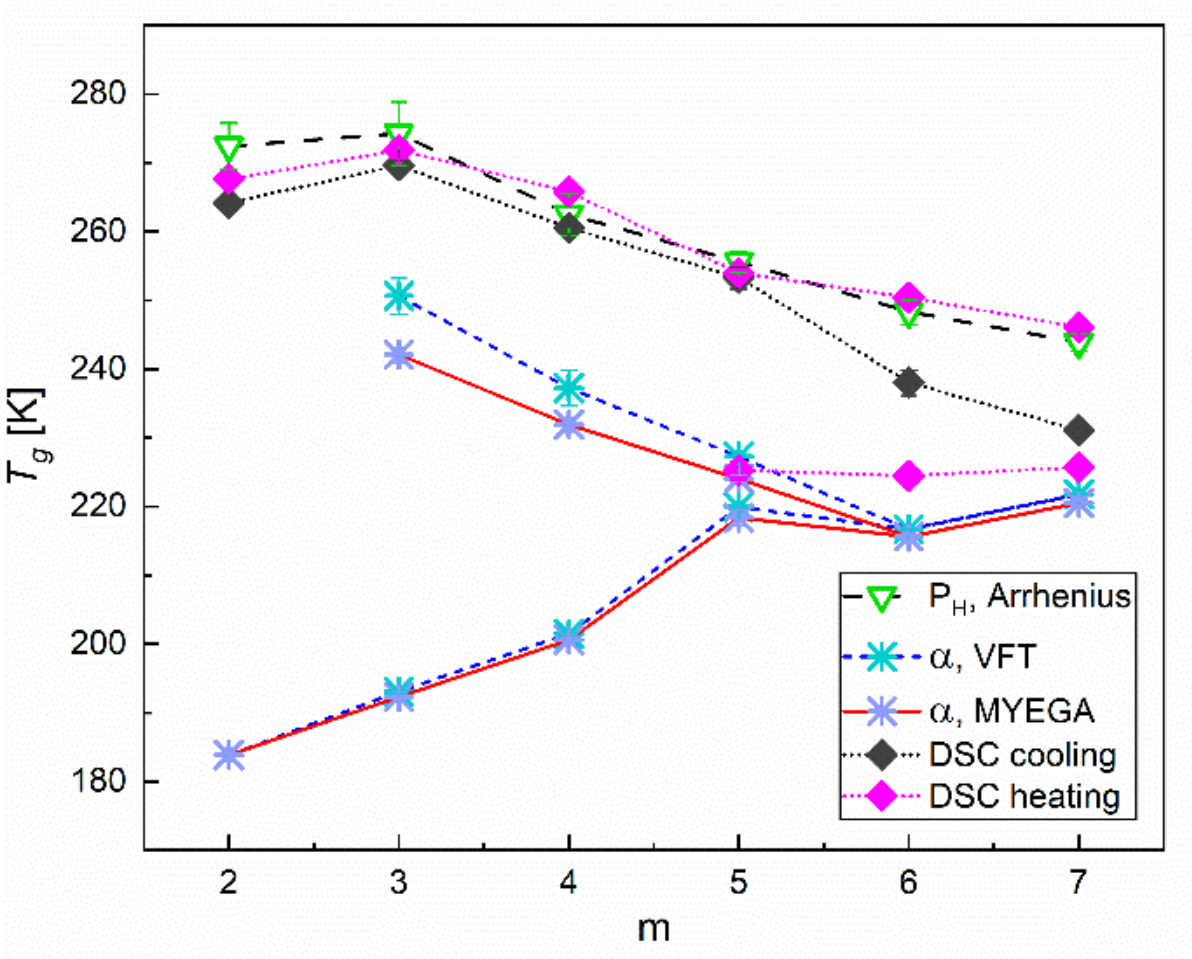

Figure 7. Calorimetric glass transition temperatures in MIXmHH6 compared to extrapolation of $P_H$ and α-relaxation times to 100 s.

### 3.3. Relaxation processes in the hexatic SmX$_A$* phase

The SmC$_A$*/SmX$_A$* transition was investigated in more detail by comparing the relaxation times of P$_{hex}$ and P$_H$ (Figure 8). For clarity, only results from cooling were analyzed, as this transition is slightly shifted on heating, and for m = 2, 3, 4 the onset of cold crystallization can additionally affect the results. The ratio of P$_{hex}$ and P$_H$ relaxation times increases on cooling, which can be described by the empirical formula:

$$\frac{\tau_{P_{hex}}}{\tau_{P_H}} = \exp\left(\left(\frac{T_0-T}{A}\right)^c\right). \tag{6}$$

The fitting parameters of Equation (6) are collected in Table S4. The power $c$ is equal to 0.32 for m = 2 and 0.13-0.18 for m = 3-7. The parameter $c$ controls the curvature of the temperature dependence, although the rate of increase of the $\tau_{P_{hex}}/\tau_{P_H}$ ratio is determined jointly by $c$ and the scaling factor $A$. Smaller $c$ means a less rapid increase of the $\tau$ ratio deep in the SmX$_A$* phase. The parameter $A$ has a unit of temperature (the ratio increases to $e$ at $T = T_0 - A$; the uncertainties are significant because of a large slope just below $T_0$). The $T_0$ temperature, corresponding to the equality of P$_{hex}$ and P$_H$ relaxation times, is the same as the onset temperature of the SmC$_A$* → SmX$_A$* transition determined by DSC for m = 3 and falls between the onset and peak temperatures in the DSC thermogram for other mixtures, see the inset in Figure 8. Equation (6) can be transformed to the form $\ln A + \ln(\ln(\tau_{P_{hex}}/\tau_{P_H}))/c = \ln(T_0 - T)$, which enables plotting all data on one line with a formula $y = x$ (Figure 9). The common scaling relationship suggests a systematic correlation between the temperature dependences of P$_{hex}$ and P$_H$ relaxation times across all mixtures, consistent with possible coupling between bond-orientational and anti-phase tilt fluctuations.

There are other literature examples of P$_{hex}$ and P$_H$ relaxation times in the SmX$_A$* phase that can be described by the empirical Equation (6) proposed herein. Based on reported data, the fitting parameters are $A$ = 2.2(1.4)·10$^{-4}$ K, $c$ = 0.14(1), $T_0$ = 307.4(1) K for the ternary mixture MIX23HH6 in 291-307 K [19], $A$ = 0.2(1.0)·10$^{-4}$ K, $c$ = 0.12(6), $T_0$ = 275(2) K for the 3F6Bi compound in 263-275 K [20], and $A$ = 2.2(1.3)·10$^{-4}$ K, $c$ = 0.15(1), $T_0$ = 357.1(7) K for the MHPSBO10 compound in 311-357 K [21]. The results from Refs. [19,20] deviate from Equation (6) at lower temperatures, probably because of partial crystallization. This analysis cannot be applied to 3F5HPhH6, 3F7HPhH6, 3F5HPhH7, where P$_{hex}$ was registered only in a single spectrum [16-18], nor to MHPNBC, MHPB(F)PBC, MHPB(H)PBC, MHPOBC, where a separate P$_{hex}$ process is not observed [13,14,15,22]. Based on available results, no obvious correlation was found between the molecular structures and the presence of behavior described by Equation (6). Thus, the dynamics in the SmX$_A$* phase may depend on various factors.

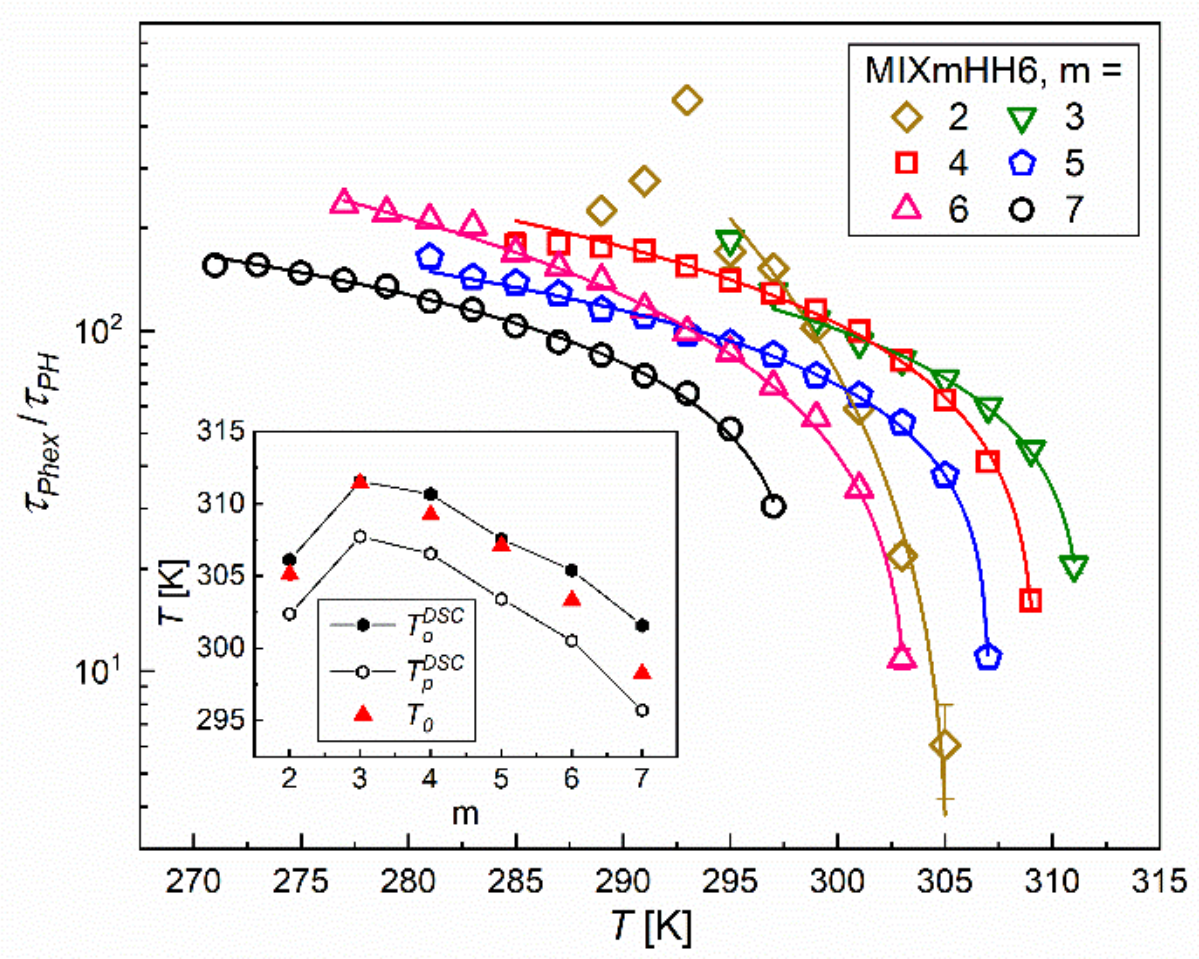

Figure 8. Ratio of relaxation times of $P_{hex}$ and $P_H$ processes in MIXmHH6 on cooling with fitting results of Equation (6). The inset shows the comparison of $T_0$ and DSC results.

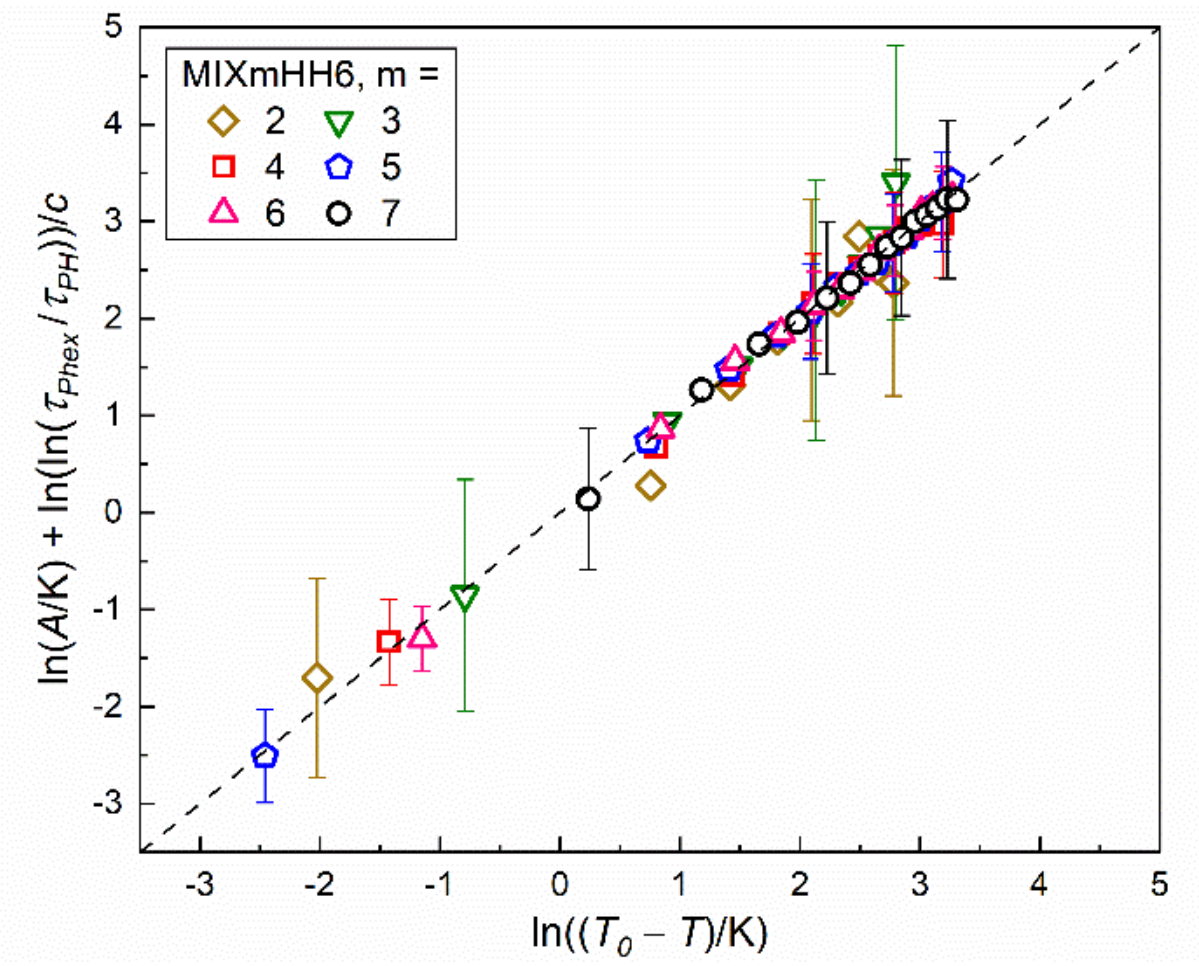

Figure 9. Ratio of relaxation times of $P_{hex}$ and $P_H$ processes scaled to a common line with a slope equal to 1 and zero intercept. The uncertainty bars are shown for selected points only.

## 5. Summary and conclusions

The $C_mH_{2m}$ chain length in the 3FmHPhH6 component of equimolar MHPOBC: 3FmHPhH6 mixtures does not influence the sequence of smectic phases nor the occurrence of a common empirical scaling relationship between the relaxation times of BOO phason $P_{hex}$ and anti-phase tilt fluctuations $P_H$ in the hexatic anticlinic $SmX_A$* phase. Meanwhile, varying m leads to an odd-even effect in the SmC*/$SmC_A$* phase transition and changes the calorimetric glass transition temperature, with a peak $T_g$ for m = 3. All mixtures are glassformers, and their tendency toward the cold crystallization on heating decreases with increasing m. The calorimetric $T_g$ correlates with collective and structural relaxation dynamics. It agrees with the extrapolation of $P_H$ relaxation time to 100 s, although for a two-step glass transition observed for m = 5-7 on heating, the lower calorimetric $T_g$ coincides quite well with the α-relaxation time extrapolated to 100 s. These findings suggest that the slowing down of both $P_H$ and α-relaxation processes may be associated with the calorimetric glass-transition behavior, with their relative importance depending on the mixture composition. The ratio of the $P_{hex}$ and $P_H$ relaxation times in all mixtures is described by the function $\exp(((T_0 - T)/A)^c)$, where $T_0$ falls in the $SmC_A$*/$SmX_A$* transition region observed by DSC. The same formula is applicable to some literature examples, although not for all of them, which means that this is not a universal behavior. The observed scaling behavior provides a basis for further investigations of the relationship between collective tilt fluctuations and bond-orientational dynamics in hexatic smectic phases.

**Acknowledgement:** Aleksandra Deptuch acknowledges the National Science Centre, Poland (grant MINIATURA 7 no. 2023/07/X/ST3/00182) for financial support.

**Conflicts of interest statement:** There are no conflicts to declare.

**Authors' contributions:**
A. Deptuch – conceptualization, investigation, formal analysis, funding acquisition, writing – original draft
M. Pączek – investigation, formal analysis, writing – review and editing
A. Drzewicz – investigation, writing – review and editing
M. Piwowarczyk – investigation, writing – review and editing
M. Urbańska – resources, writing – review and editing
E. Juszyńska-Gałązka – investigation, writing – review and editing

# Collective relaxation dynamics and vitrification in hexatic smectic binary equimolar mixtures containing component with varying $C_3F_7CH_2OC_mH_{2m}$- chain length

Aleksandra Deptuch[1,*], Mateusz Pączek[2], Anna Drzewicz[1], Marcin Piwowarczyk[1], Magdalena Urbańska[3], Ewa Juszyńska-Gałązka[1,4]

[1] The Henryk Niewodniczański Institute of Nuclear Physics Polish Academy of Sciences, Radzikowskiego 152, PL-31342 Kraków, Poland

[2] Faculty of Chemistry, Jagiellonian University, Gronostajowa 2, PL-30387 Kraków, Poland

[3] Institute of Chemistry, Military University of Technology, Kaliskiego 2, PL-00908 Warsaw, Poland

[4] Research Center for Thermal and Entropic Science, Graduate School of Science, Osaka University, 560-0043 Osaka, Japan

*corresponding author, aleksandra.deptuch@ifj.edu.pl

## Supplementary Materials

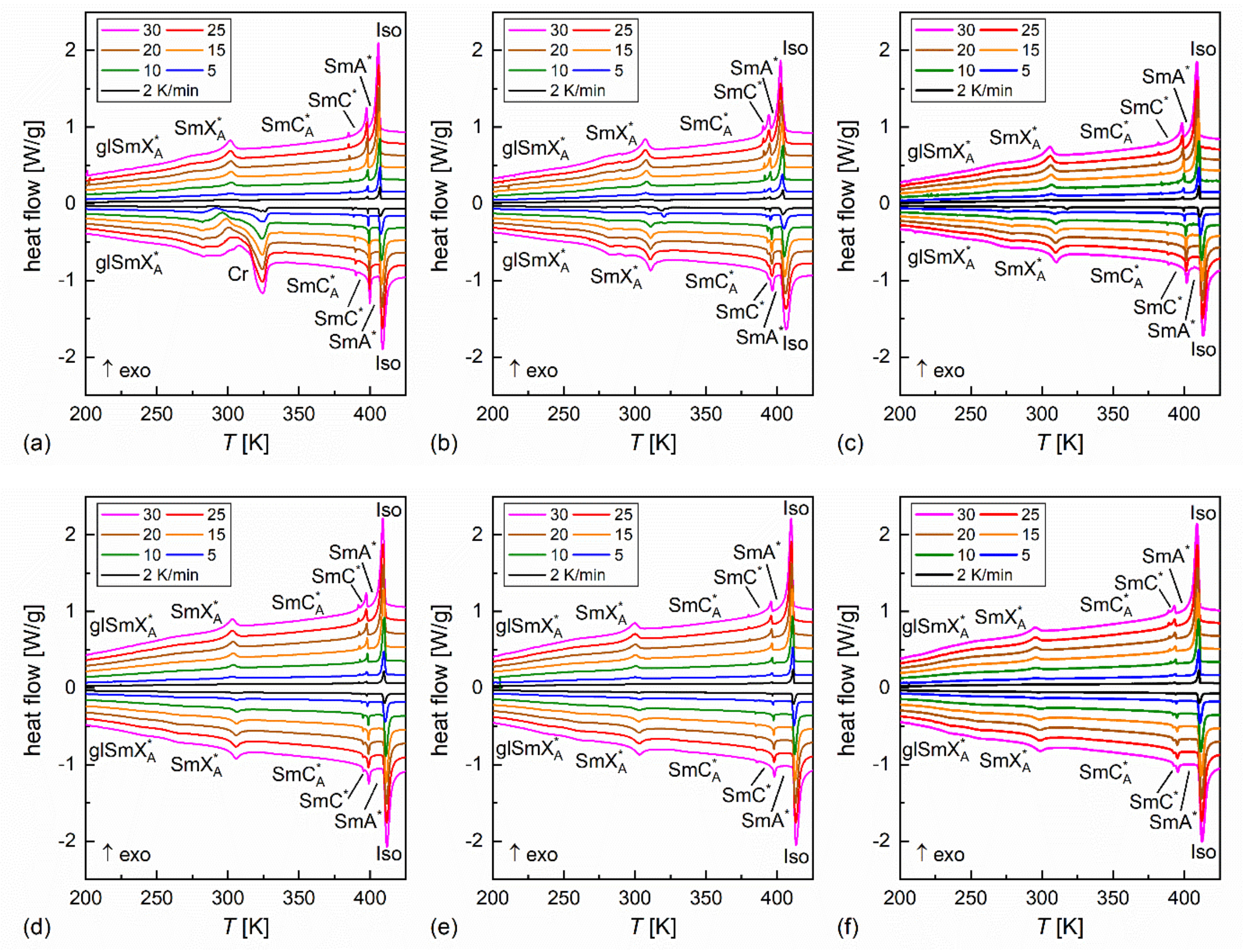

Figure S1. DSC thermograms of MIXmHH6 with m = 2 (a), 3 (b), 4 (c), 5 (d), 6 (e), 7 (f).

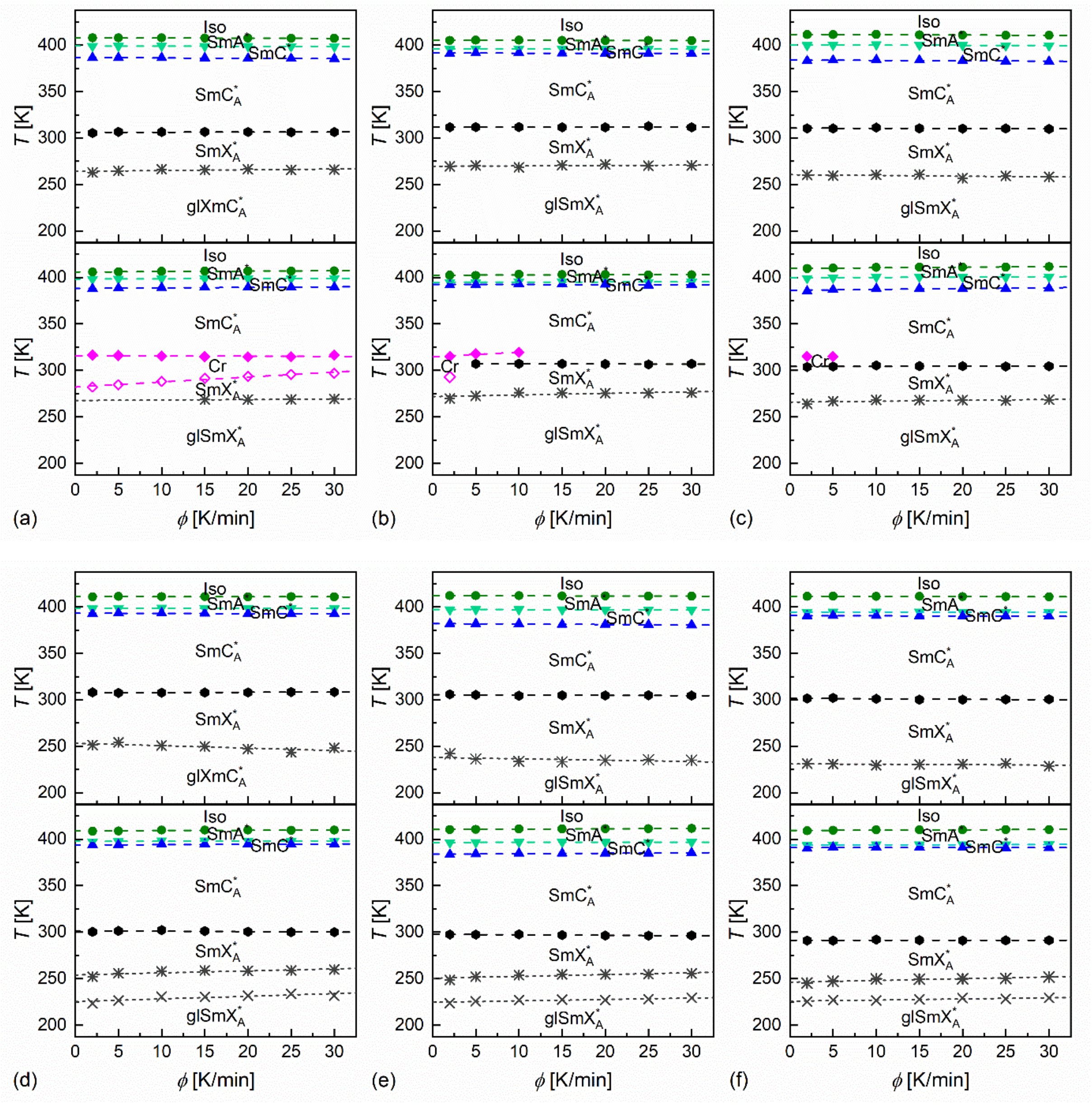

Figure S2. Phase transition temperatures vs. the cooling/heating rate for MIXmHH6 with m = 2 (a), 3 (b), 4 (c), 5 (d), 6 (e), 7 (f). Upper and bottom panels show the results obtained during cooling and heating, respectively.

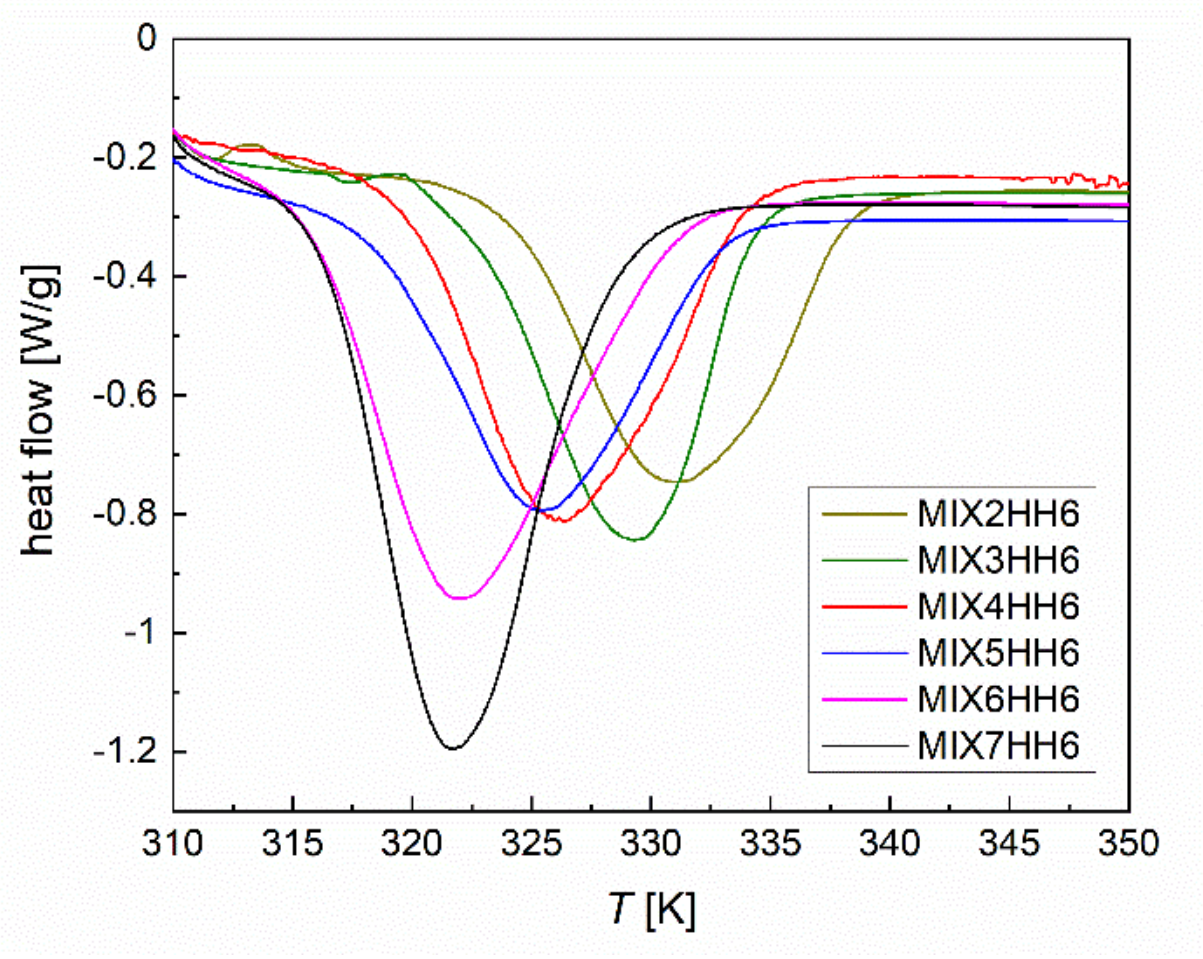

Figure S3. Melting endotherms of MIXmHH6 during the 1st heating at 10 K/min.

Table S1. Phase transition temperatures in MIXmHH6 (1st row, extrapolated to 0 K/min) with the corresponding enthalpy changes (*2rd row*) determined by DSC on cooling/heating. The temperatures are given in K and the enthalpy changes are given in kJ/mol. Only the values for melting (Cr/SmC$_A$*) are based on the DSC thermograms from the 1st heating at 10 K/min.

| mixture | Cr/SmC$_A$* | SmX$_A$*/SmC$_A$* | SmC$_A$*/SmC* | SmC*/SmA* | SmA*/Iso |
|---|---|---|---|---|---|
| MIX2HH6 | 324.0<br>*18.1* | 306.1 / -<br>*1.6 / -* | 386.6 / 388.1<br>*0.1 / 0.1* | 399.0 / 398.3<br>*1.5 / 1.6* | 407.8 / 405.9<br>*5.6 / 5.5* |
| MIX3HH6 | 322.3<br>*17.8* | 311.6 / 289.5<br>*1.8 / 1.6* | 391.4 / 392.3<br>*0.5 / 0.5* | 395.9 / 394.8<br>*1.1 / 1.2* | 405.2 / 402.5<br>*6.0 / 6.0* |
| MIX4HH6 | 319.7<br>*22.5* | 310.7 / 304.2<br>*1.9 / 1.6* | 383.9 / 386.1<br>*0.1 / 0.1* | 400.2 / 399.6<br>*1.5 / 1.4* | 411.4 / 409.8<br>*6.3 / 6.1* |
| MIX5HH6 | 318.3<br>*18.6* | 307.6 / 301.4<br>*1.8 / 1.6* | 393.1 / 393.8<br>*0.3 / 0.4* | 398.3 / 397.7<br>*0.9 / 1.1* | 411.0 / 409.0<br>*6.6 / 6.6* |
| MIX6HH6 | 315.6<br>*25.9* | 305.4 / 297.6<br>*1.6 / 1.4* | 382.0 / 383.8<br>*0.1 / 0.1* | 397.2 / 396.6<br>*1.1 / 1.2* | 412.0 / 410.5<br>*7.1 / 6.8* |
| MIX7HH6 | 316.2<br>*29.4* | 301.6 / 291.2<br>*1.2 / 1.3* | 390.6 / 390.8<br>*0.4 / 0.3* | 394.4 / 393.7<br>*0.5 / 0.8* | 411.3 / 409.3<br>*7.2 / 6.9* |

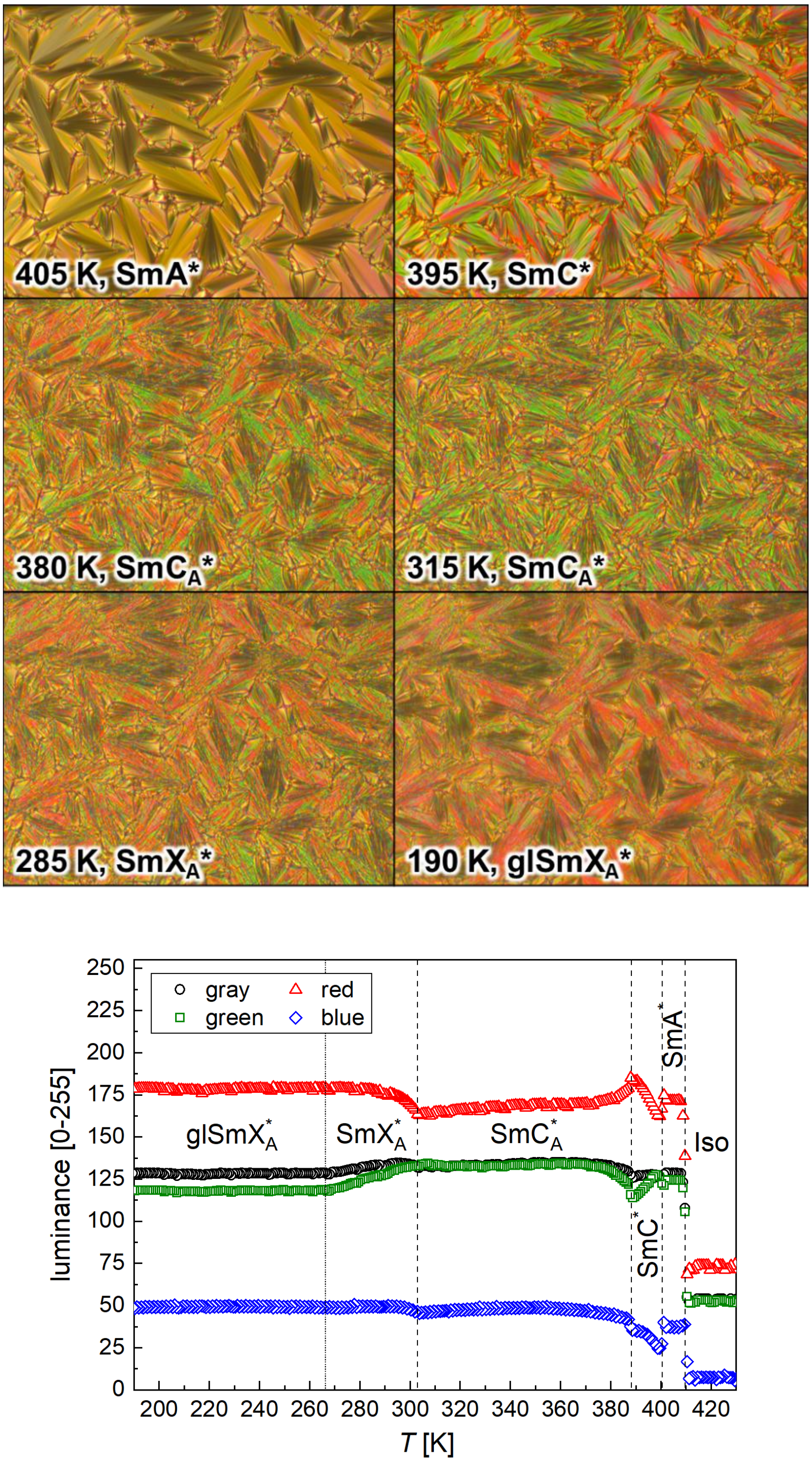

Figure S4. Selected POM textures of MIX2HH6 collected in a transmission mode during cooling at 10 K/min. Each texture covers an area of 622 × 466 μm$^2$. The plot beneath shows the average luminance each texture (red, green, blue, and weighted gray). The glass transition temperature is based on DSC results.

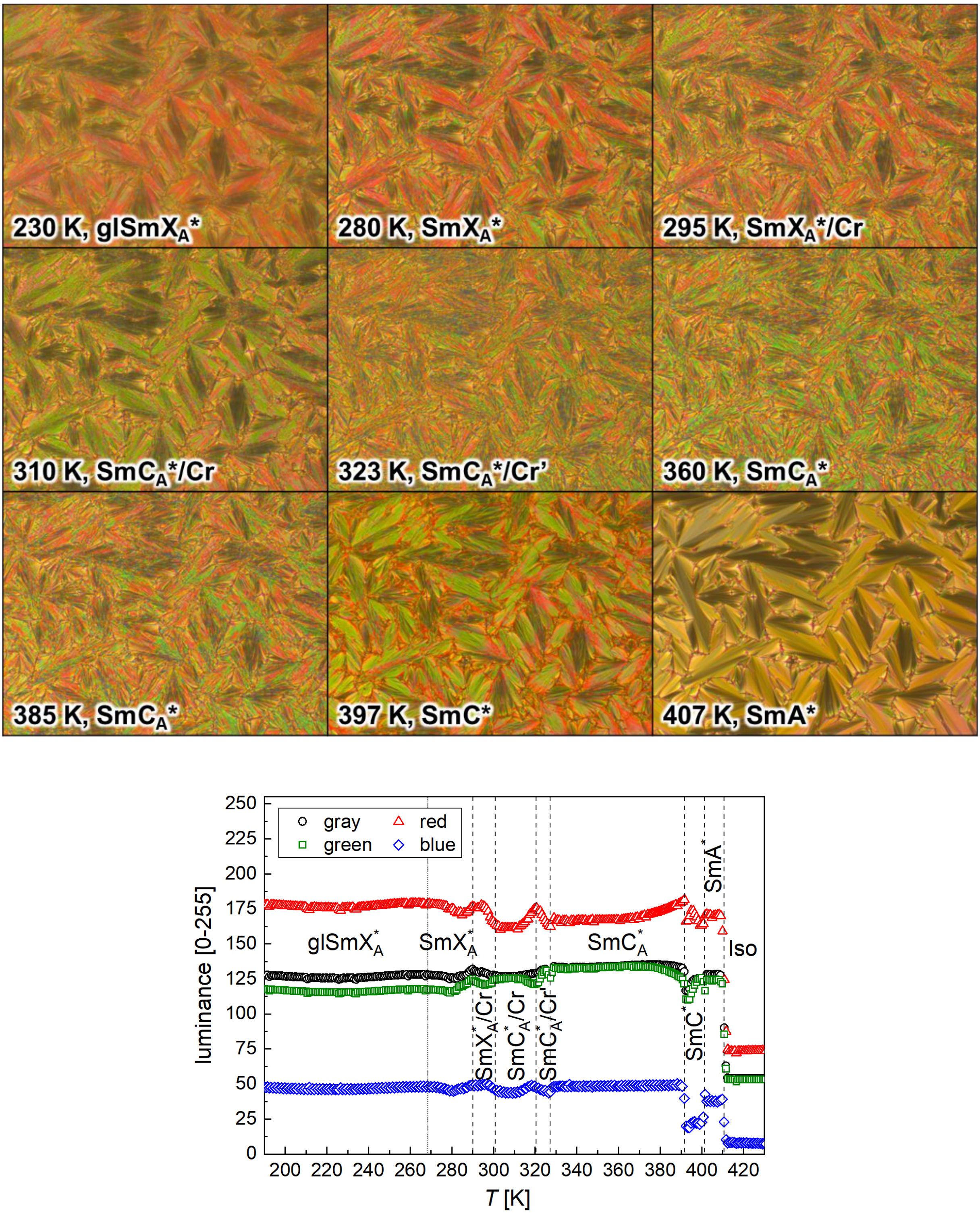

Figure S5. Selected POM textures of MIX2HH6 collected in a transmission mode during heating at 10 K/min. Each texture covers an area of 622 × 466 μm$^2$. The plot beneath shows the average luminance each texture (red, green, blue, and weighted gray). The glass transition temperature is based on DSC results.

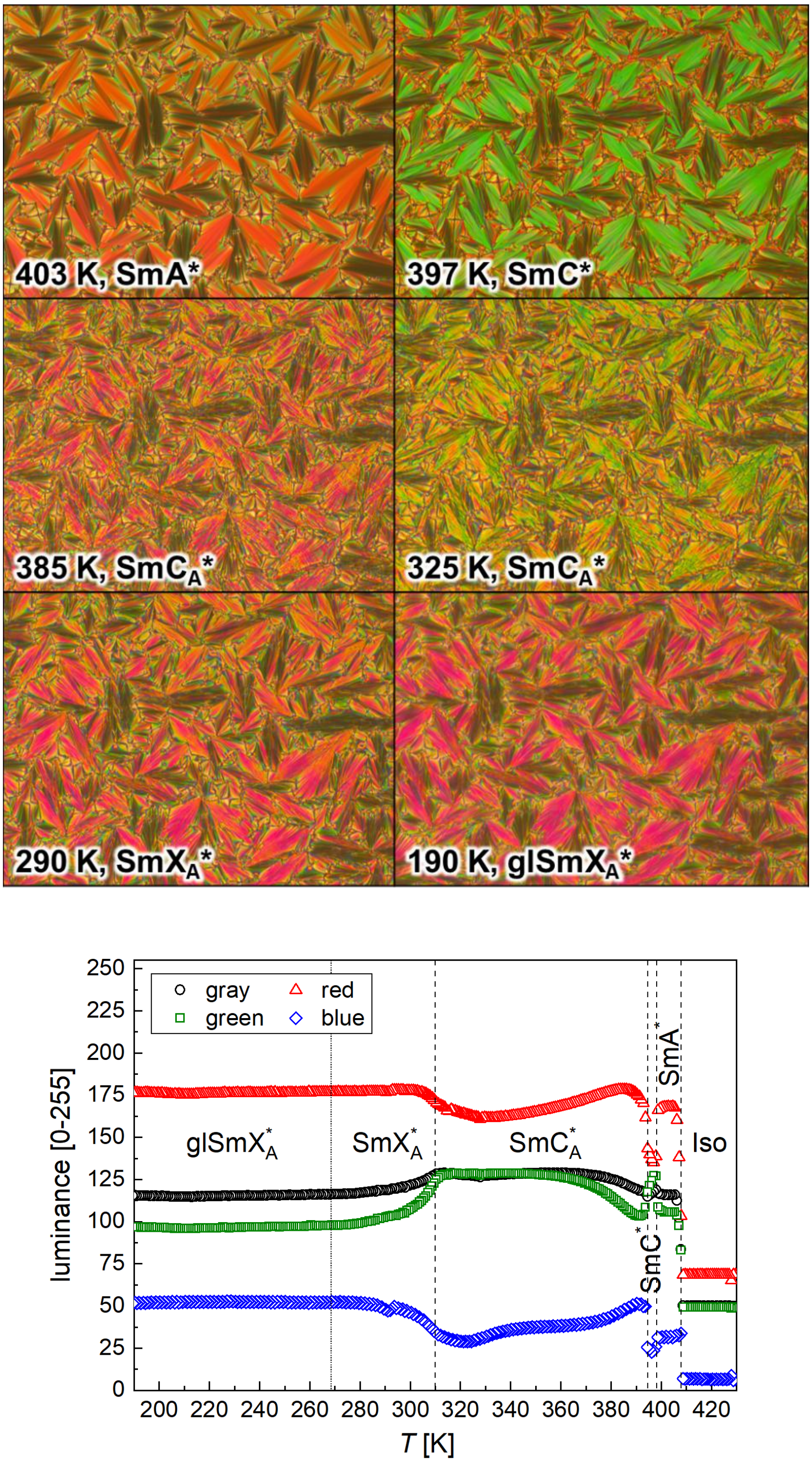

Figure S6. Selected POM textures of MIX3HH6 collected in a transmission mode during cooling at 10 K/min. Each texture covers an area of 622 × 466 μm$^2$. The plot beneath shows the average luminance of each texture (red, green, blue, and weighted gray). The glass transition temperature is based on DSC results.

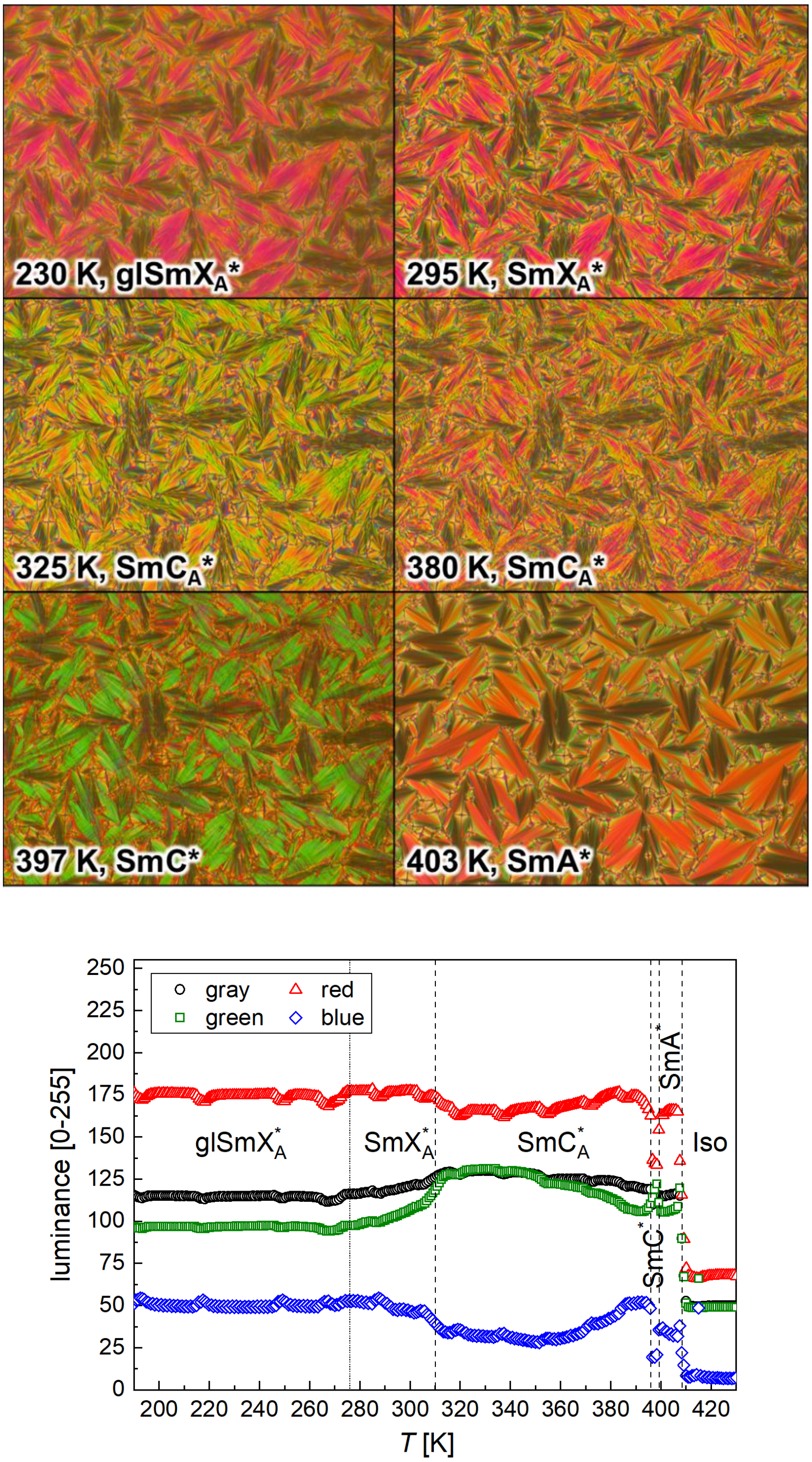

Figure S7. Selected POM textures of MIX3HH6 collected in a transmission mode during heating at 10 K/min. Each texture covers an area of 622 × 466 μm$^2$. The plot beneath shows the average luminance of each texture (red, green, blue, and weighted gray). The glass transition temperature is based on DSC results.

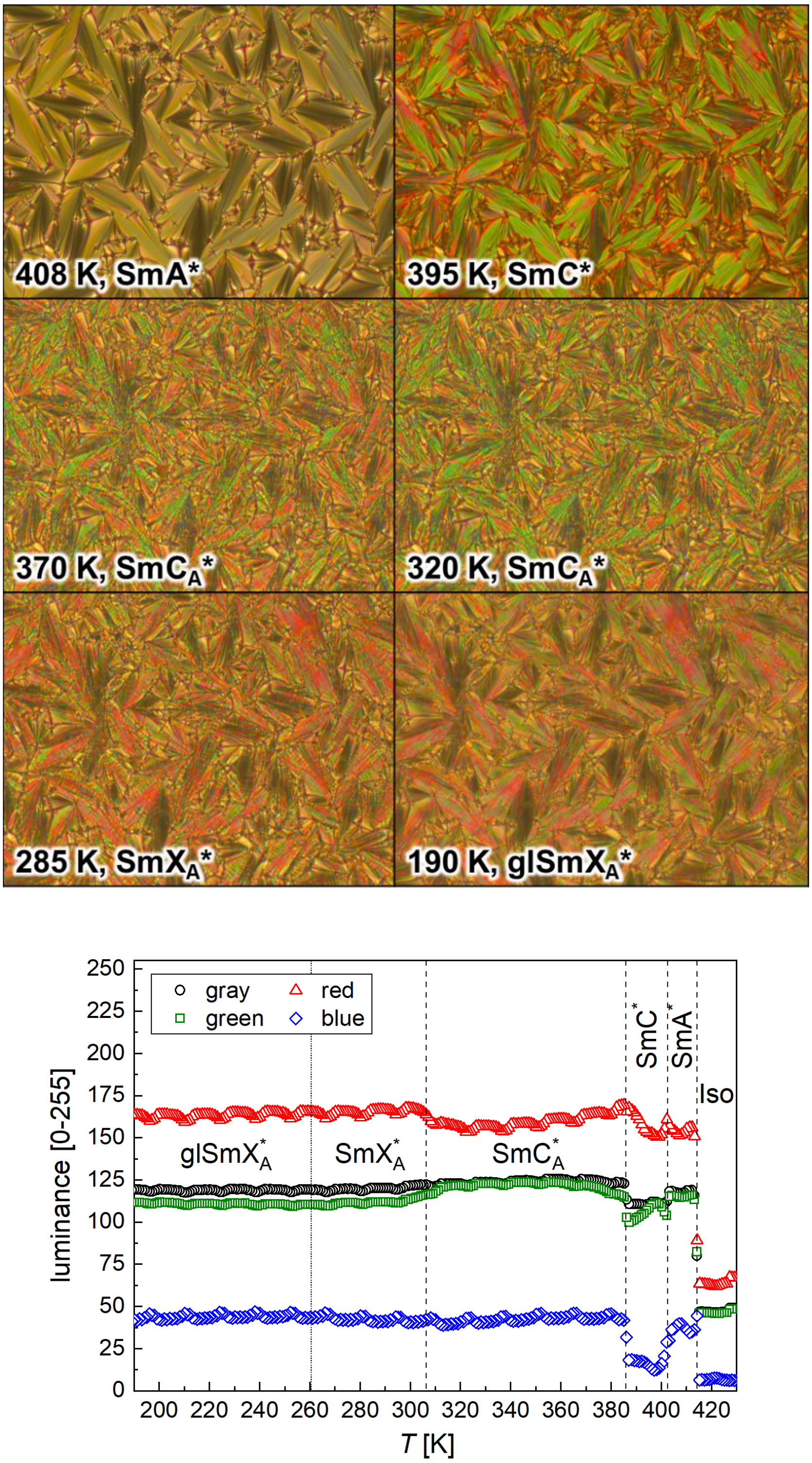

Figure S8. Selected POM textures of MIX4HH6 collected in a transmission mode during cooling at 10 K/min. Each texture covers an area of 622 × 466 μm$^2$. The plot beneath shows the average luminance of each texture (red, green, blue, and weighted gray). The glass transition temperature is based on DSC results.

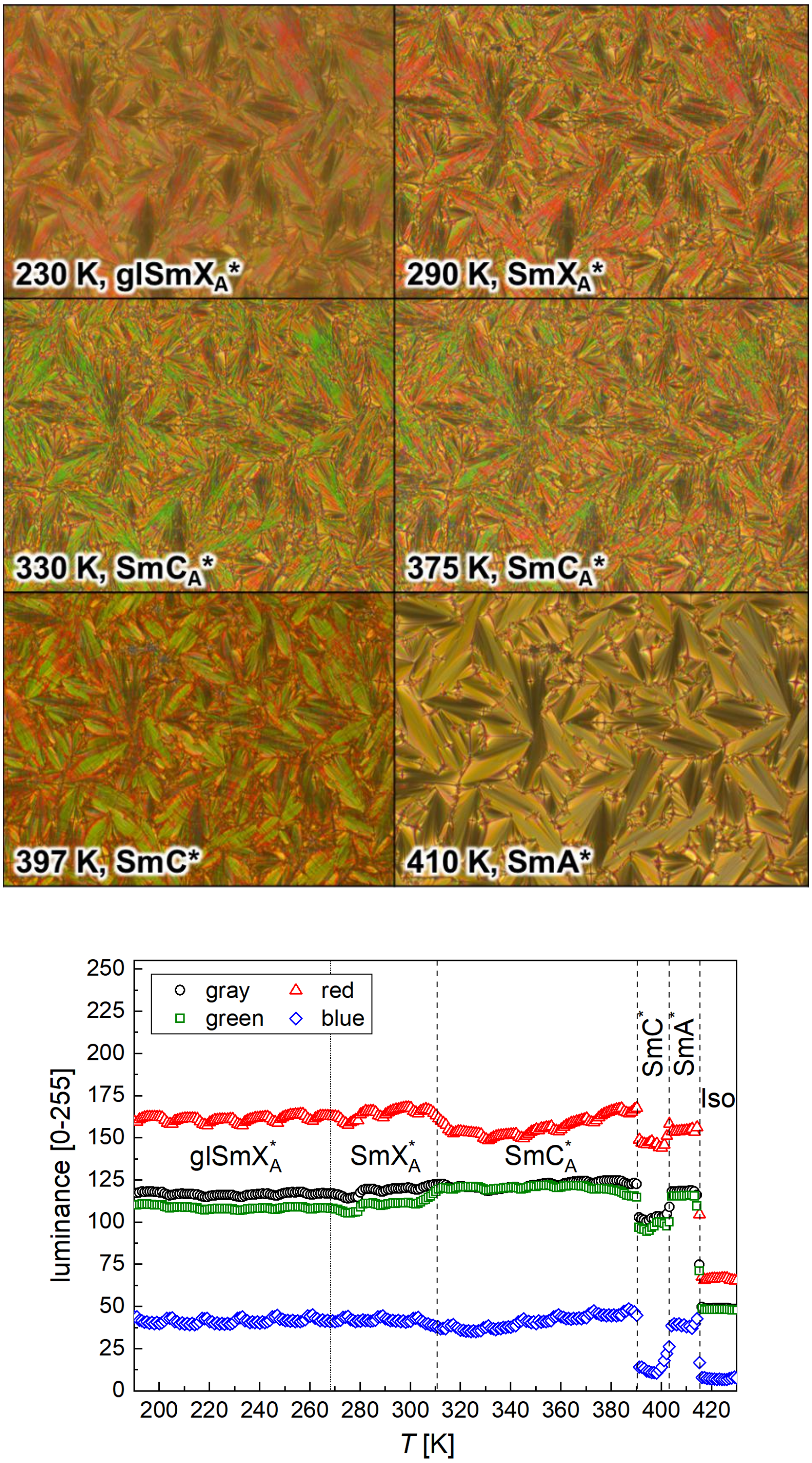

Figure S9. Selected POM textures of MIX4HH6 collected in a transmission mode during heating at 10 K/min. Each texture covers an area of 622 × 466 μm$^2$. The plot beneath shows the average luminance of each texture (red, green, blue, and weighted gray). The glass transition temperature is based on DSC results.

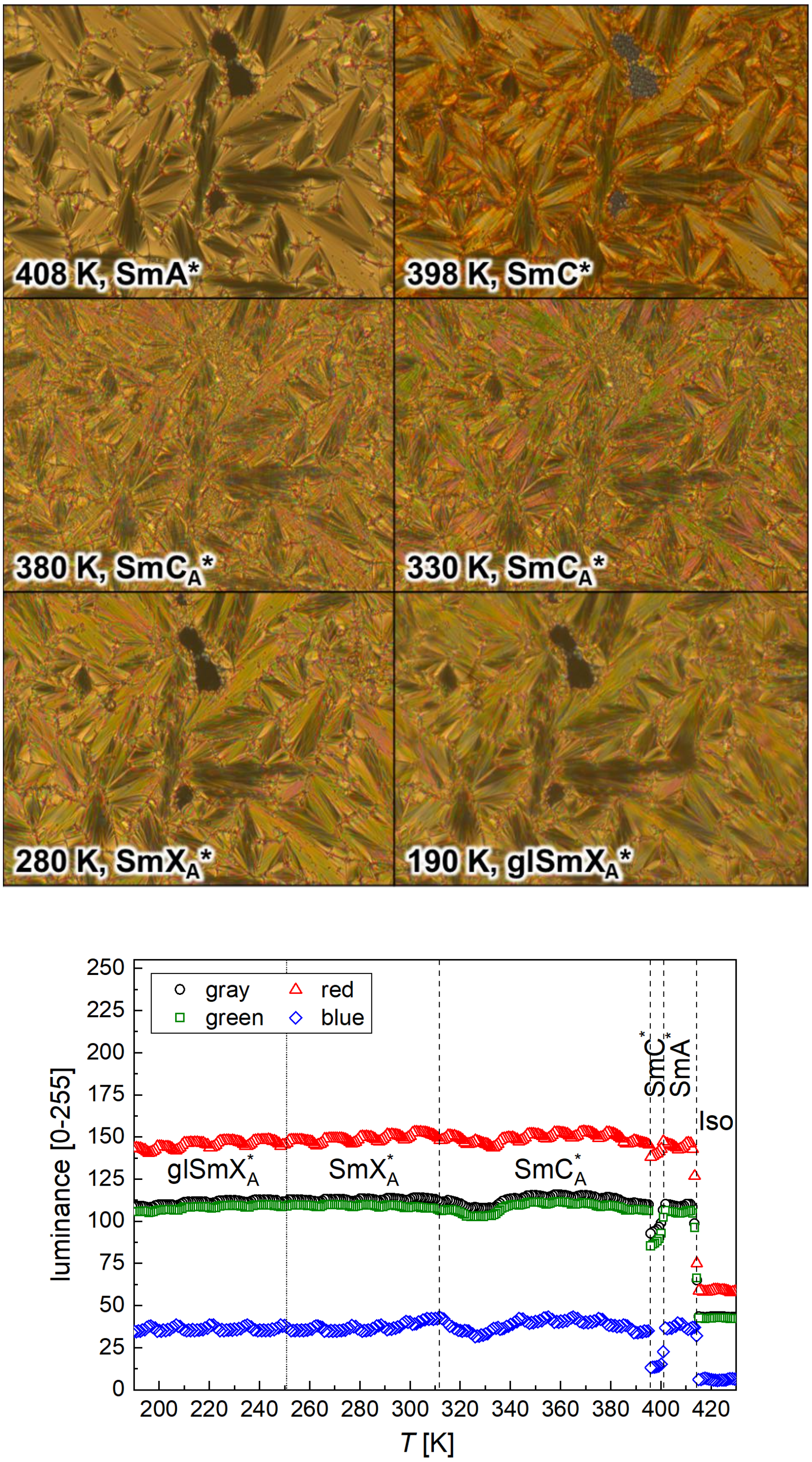

Figure S10. Selected POM textures of MIX5HH6 collected in a transmission mode during cooling at 10 K/min. Each texture covers an area of 622 × 466 μm$^2$. The plot beneath shows the average luminance of each texture (red, green, blue, and weighted gray). The SmC$_A$* → SmX$_A$* and glass transition temperatures are based on DSC results.

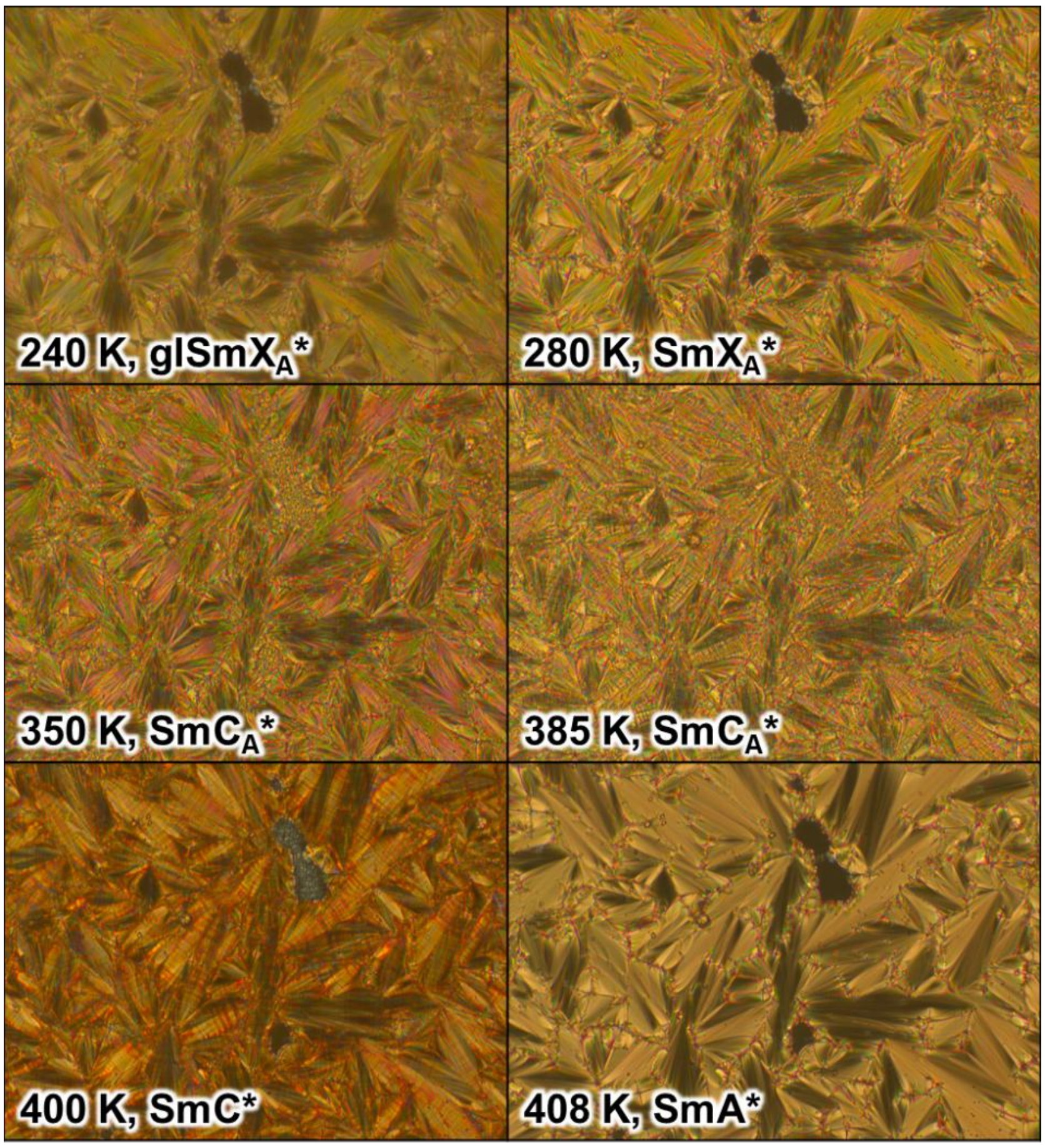

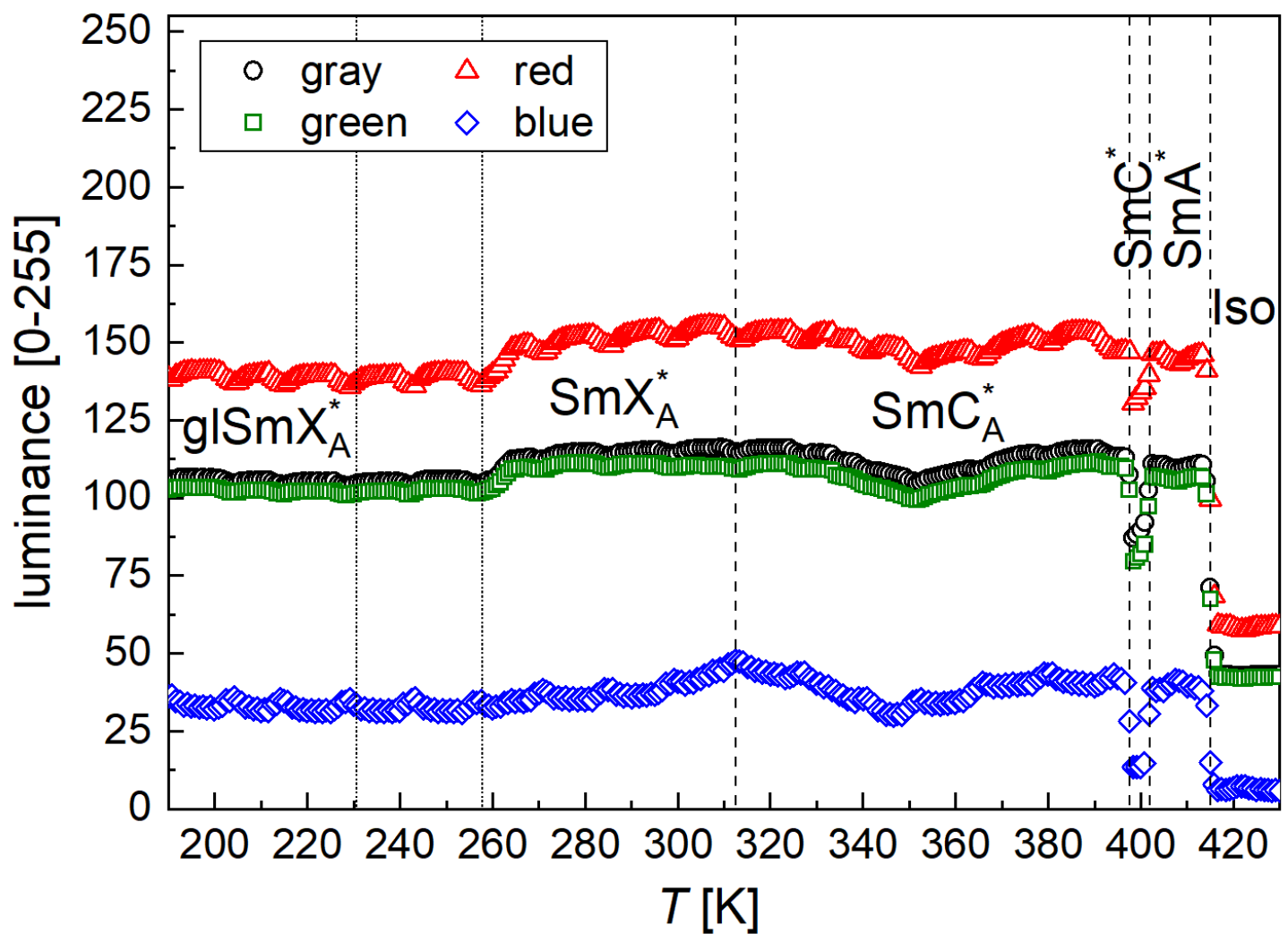

Figure S11. Selected POM textures of MIX5HH6 collected in a transmission mode during heating at 10 K/min. Each texture covers an area of 622 × 466 μm$^2$. The plot beneath shows the average luminance of each texture (red, green, blue, and weighted gray). The double glass transition and $SmX_A$* → $SmC_A$* temperatures are based on DSC results.

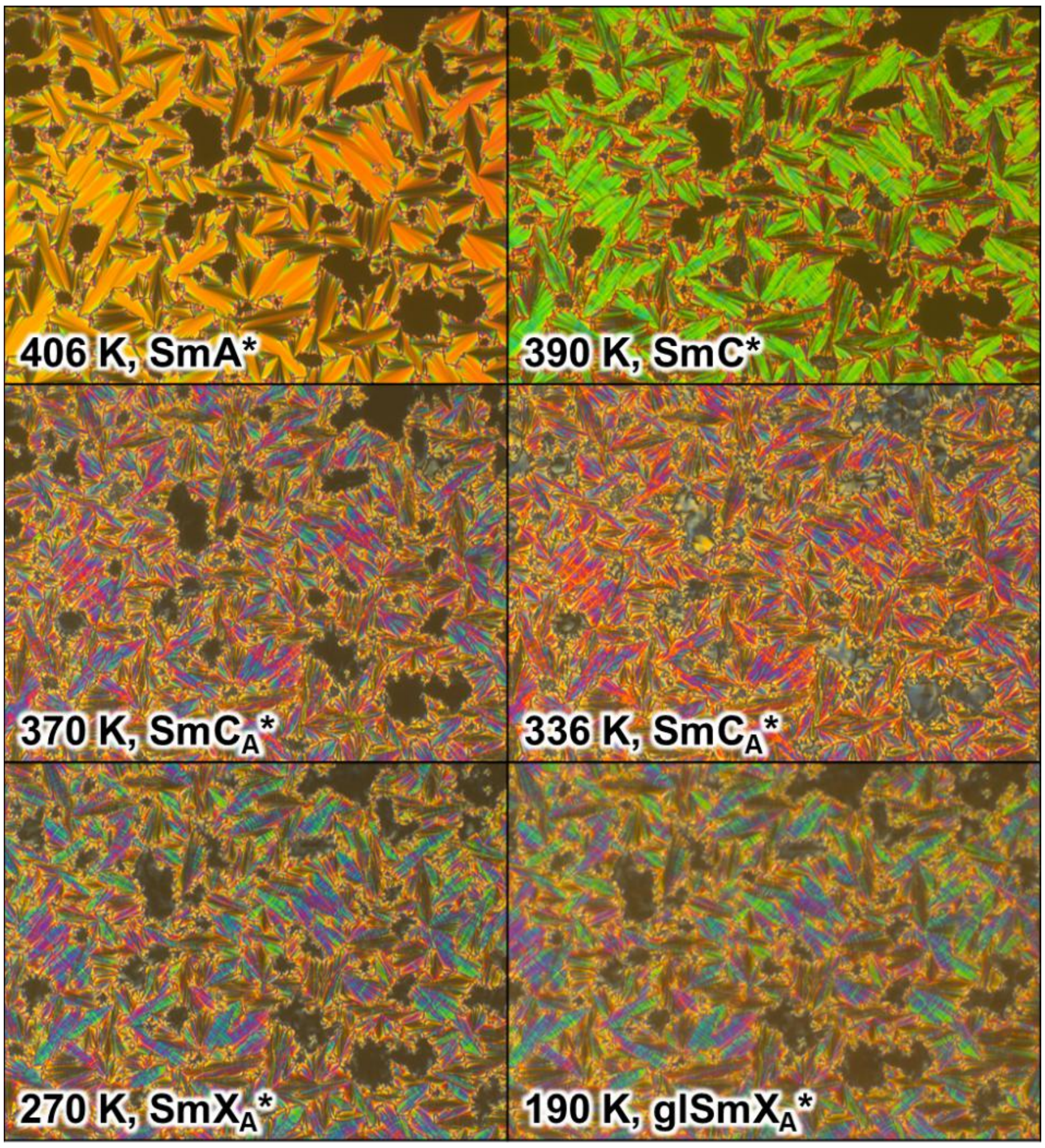

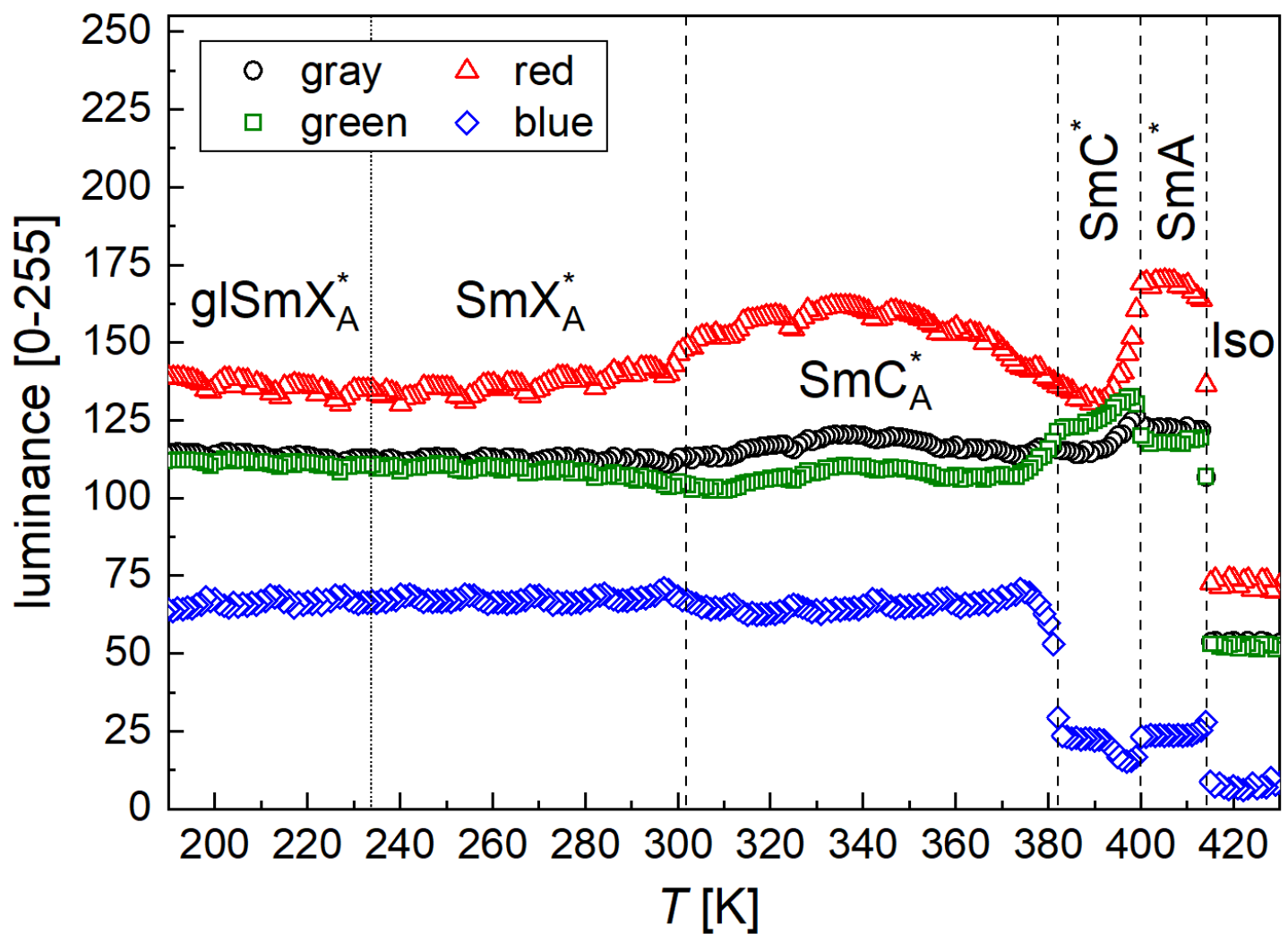

Figure S12. Selected POM textures of MIX6HH6 collected in a transmission mode during cooling at 10 K/min. Each texture covers an area of 622 × 466 μm$^2$. The plot beneath shows the average luminance of each texture (red, green, blue, and weighted gray). The double glass transition and SmX$_A$* → SmC$_A$* temperatures are based on DSC results.

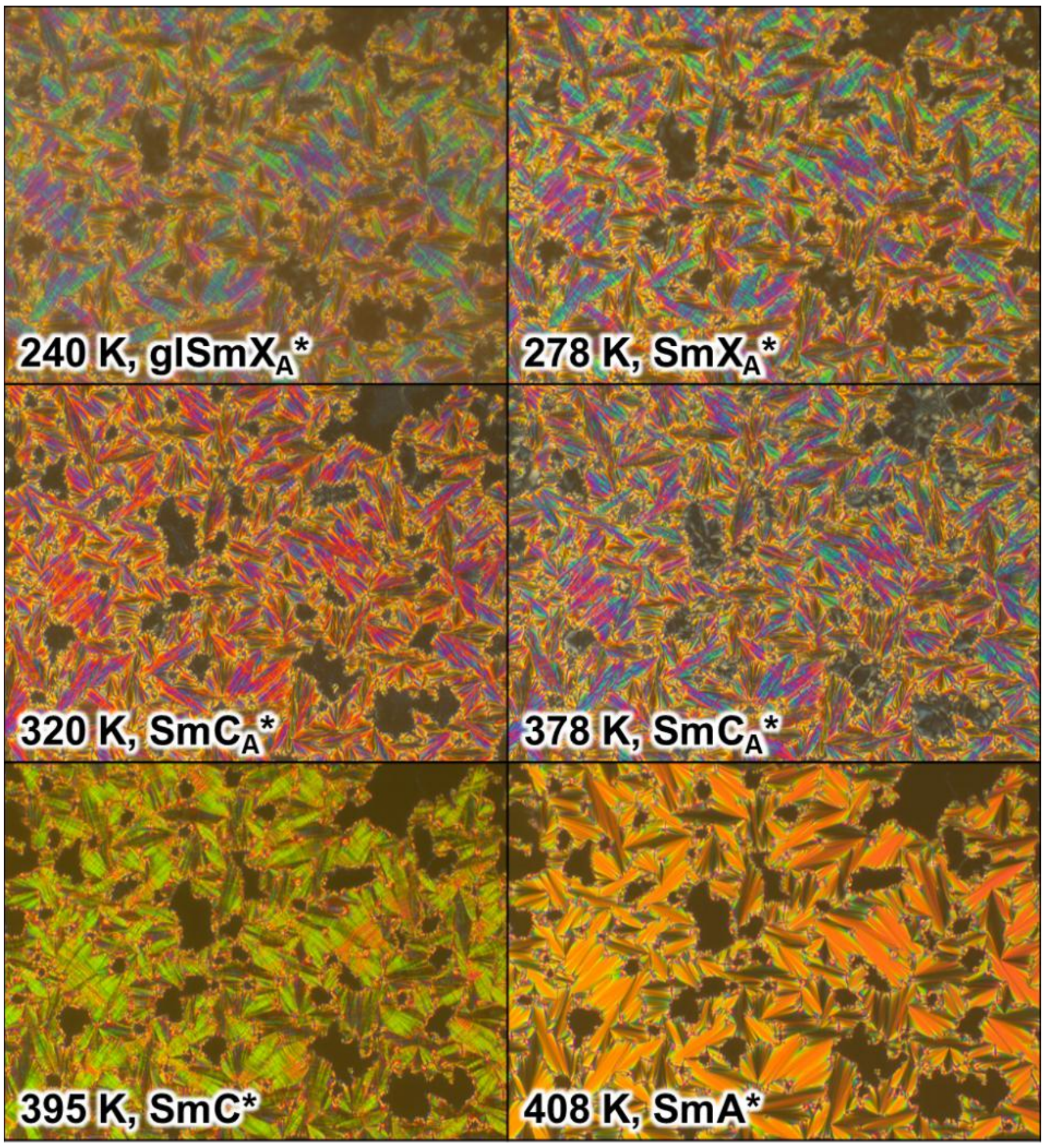

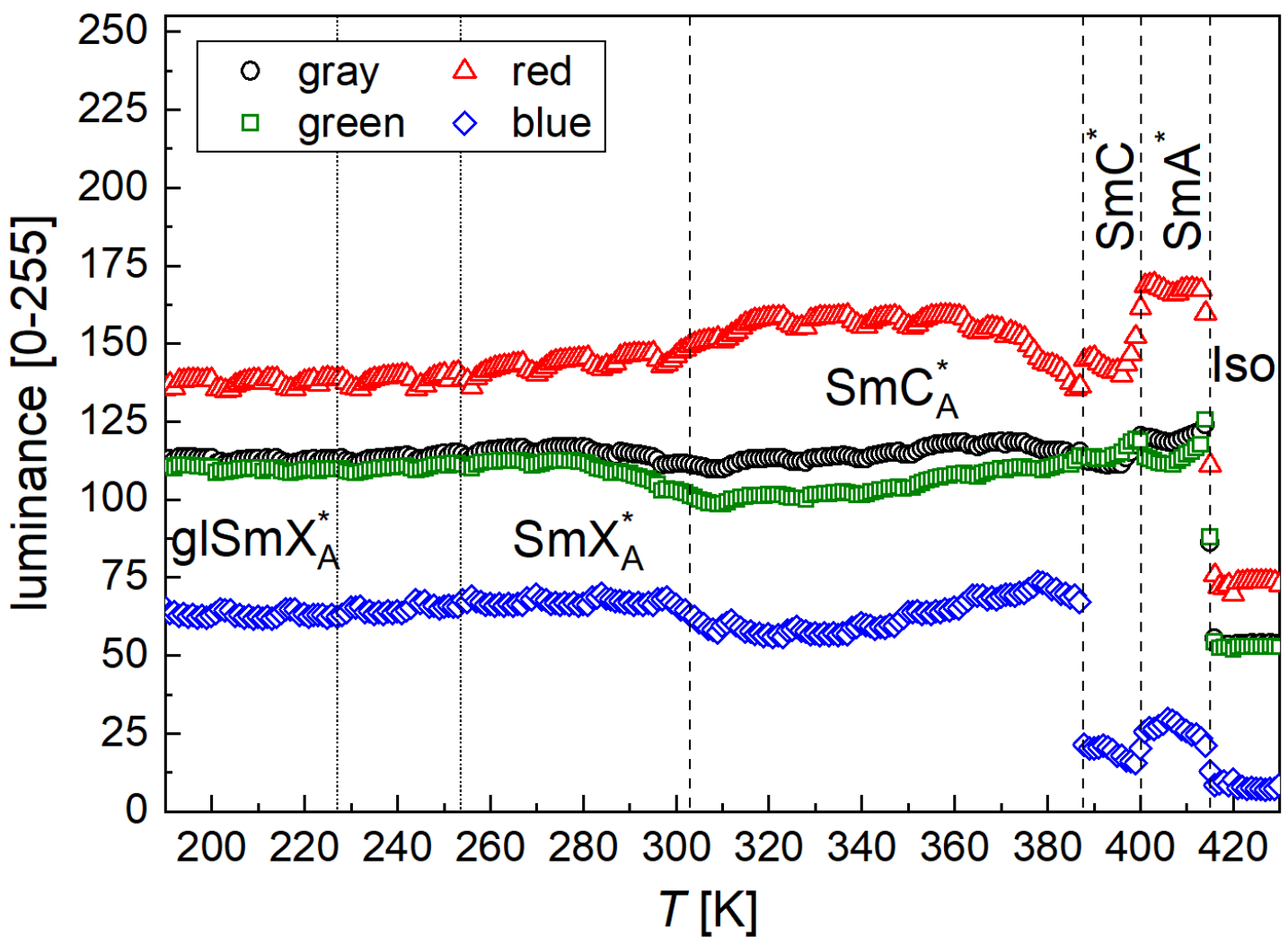

Figure S13. Selected POM textures of MIX6HH6 collected in a transmission mode during heating at 10 K/min. Each texture covers an area of 622 × 466 μm$^2$. The plot beneath shows the average luminance of each texture (red, green, blue, and weighted gray). The double glass transition and SmX$_A$* → SmC$_A$* temperatures are based on DSC results.

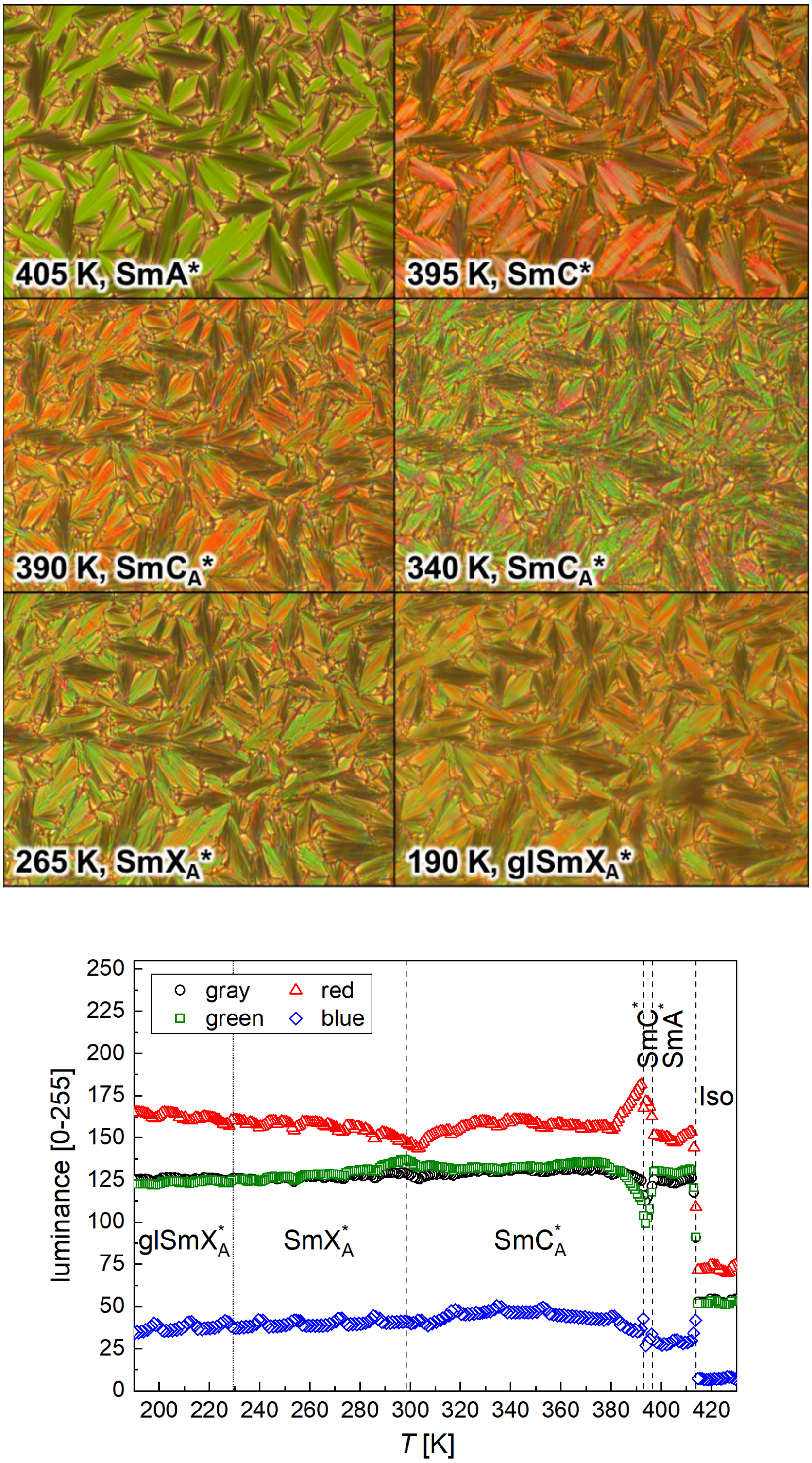

Figure S14. Selected POM textures of MIX7HH6 collected in a transmission mode during cooling at 10 K/min. Each texture covers an area of 622 × 466 μm$^2$. The plot beneath shows the average luminance of each texture (red, green, blue, and weighted gray). The double glass transition and SmX$_A$* → SmC$_A$* temperatures are based on DSC results.

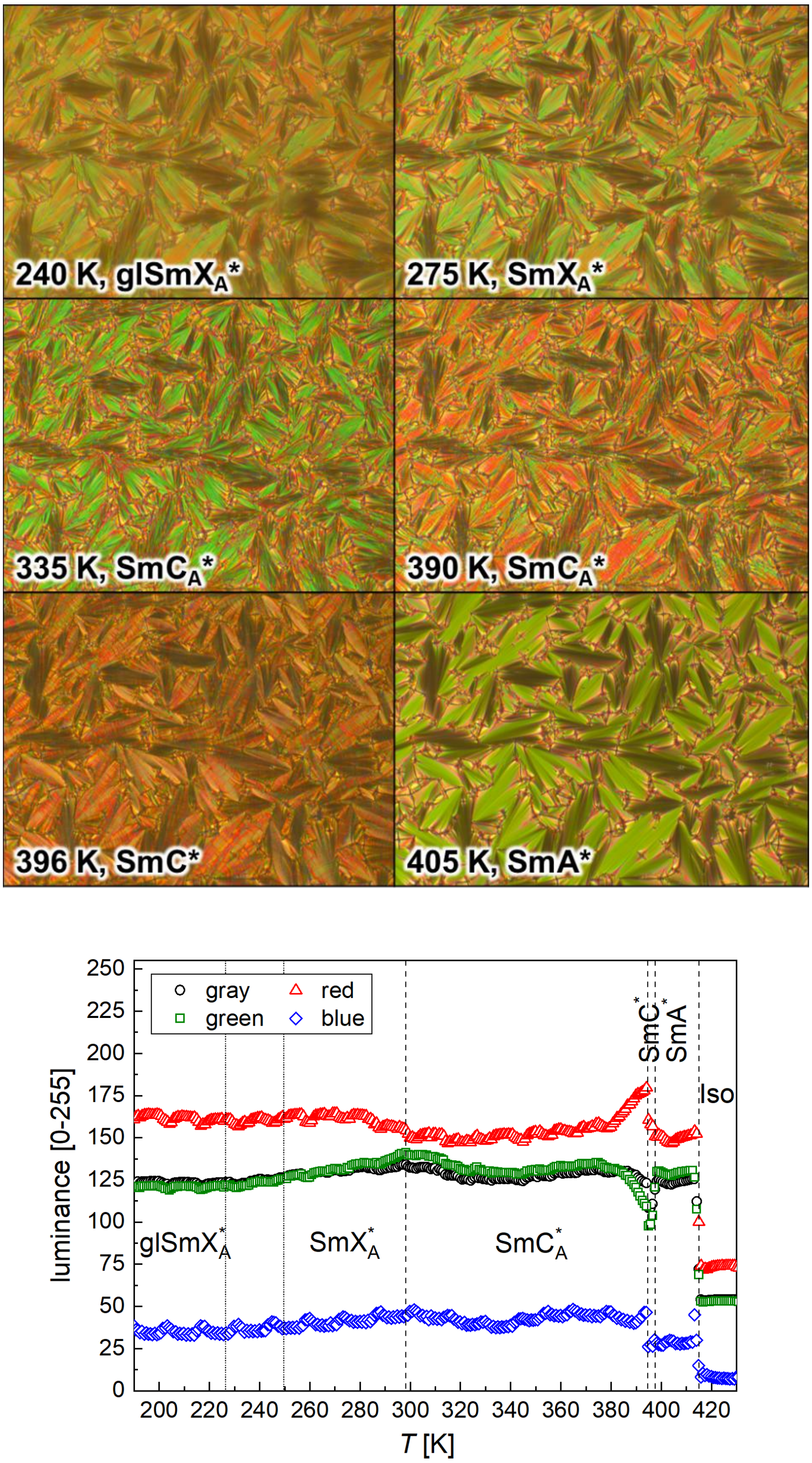

Figure S15. Selected POM textures of MIX7HH6 collected in a transmission mode during heating at 10 K/min. Each texture covers an area of 622 × 466 μm². The plot beneath shows the average luminance of each texture (red, green, blue, and weighted gray). The double glass transition and $SmX_A^* \rightarrow SmC_A^*$ temperatures are based on DSC results.

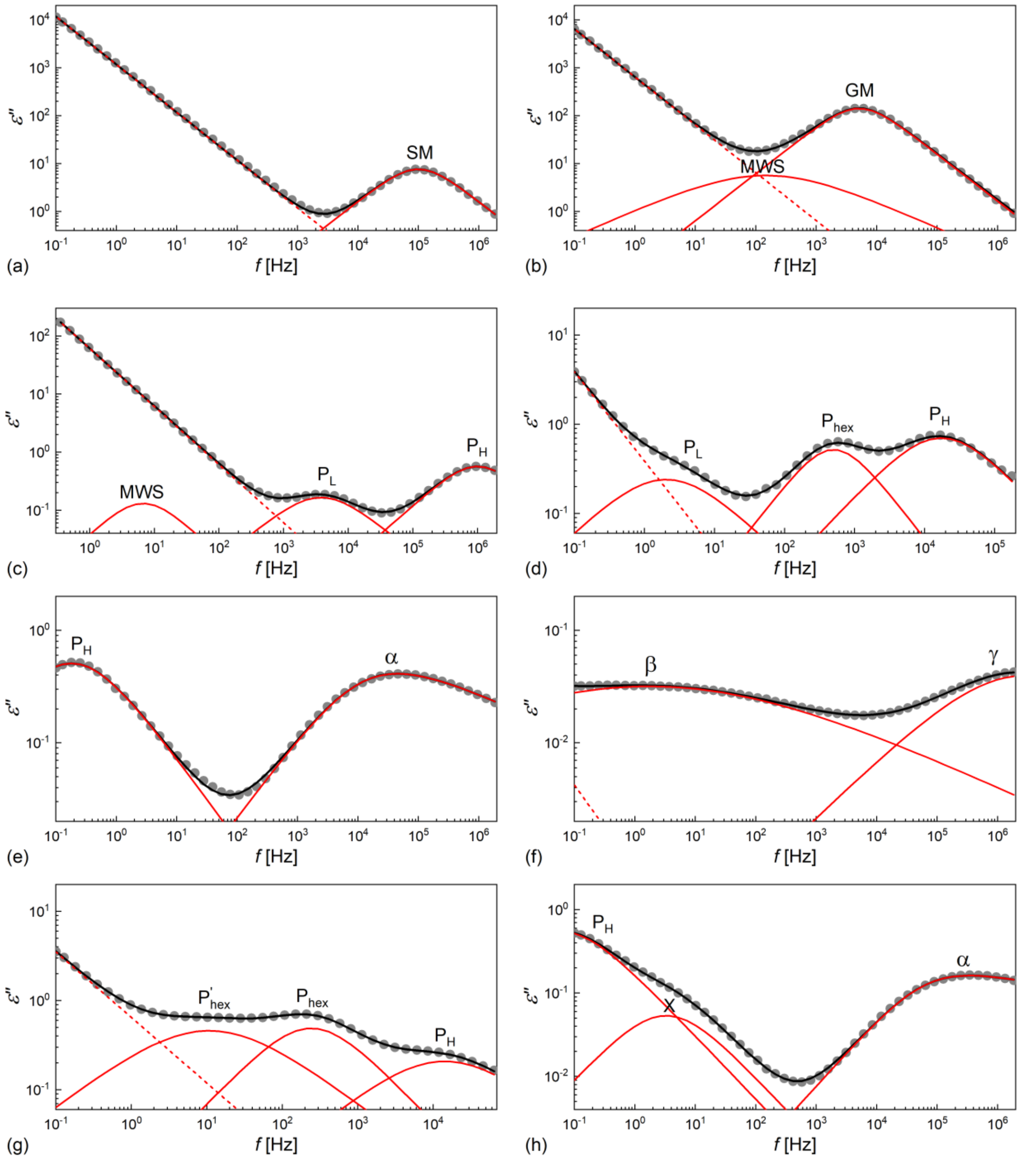

Figure S16. Fitting results of the Havriliak-Negami model (lines) to experimental BDS spectra (points) of MIX6HH6 in different phases: SmA* at 399 K (a), SmC* at 393 K (b), $SmC_A$* at 353 K (c), $SmX_A$* at 301 K (d), $SmX_A$* at 263 K (e), gl$SmX_A$* at 183 K (f); additional relaxation processes in the $SmX_A$* phase of MIX2HH6 at 301 K (g) and of MIX3HH6 at 283 K (h). All results obtained on cooling.

Table S2. Activation energies in kJ/mol of ionic conductivity and relaxation processes in MIXmHH6.

| m | 2 | 3 | 4 | 5 | 6 | 7 |
|---|---|---|---|---|---|---|
| $\sigma_{ion}$, $SmC_A$* | 54.3(3) | 60.3(2) | 54.7(3) | 70.5(2) | 62.2(1) | 56.5(3) |
| MWS, $SmC_A$* | 61.0(6) | 67.6(7) | 60(1) | 77(1) | 68(2) | 62(2) |
| $P_L$, $SmC_A$* | 110.9(5) | 115.1(4) | 110.2(6) | 114.4(5) | 107.1(5) | 105.0(4) |
| $P_H$, $SmX_A$* | 379(5) | 328(6) | 249(3) | 210(2) | 190(2) | 178(1) |
| $P_{hex}$, $SmX_A$* | 504(5) | 355(4) | 274(4) | 235(2) | 231(3) | 197(2) |
| α, $SmX_A$* | 83(2) | – [a] | – [a] | – [a] | – [a] | – [a] |
| β, gl$SmX_A$* | 48.4(4) | 43(2) | 45.6(8) | 54(2) | 48.0(9) | 65.6(6) |
| γ, gl$SmX_A$* | 17(1) | – [b] | 16.0(3) | 15(3) | 14.2(7) | 12(1) |

[a] non-Arrhenius dependence of relaxation time

[b] too few experimental points

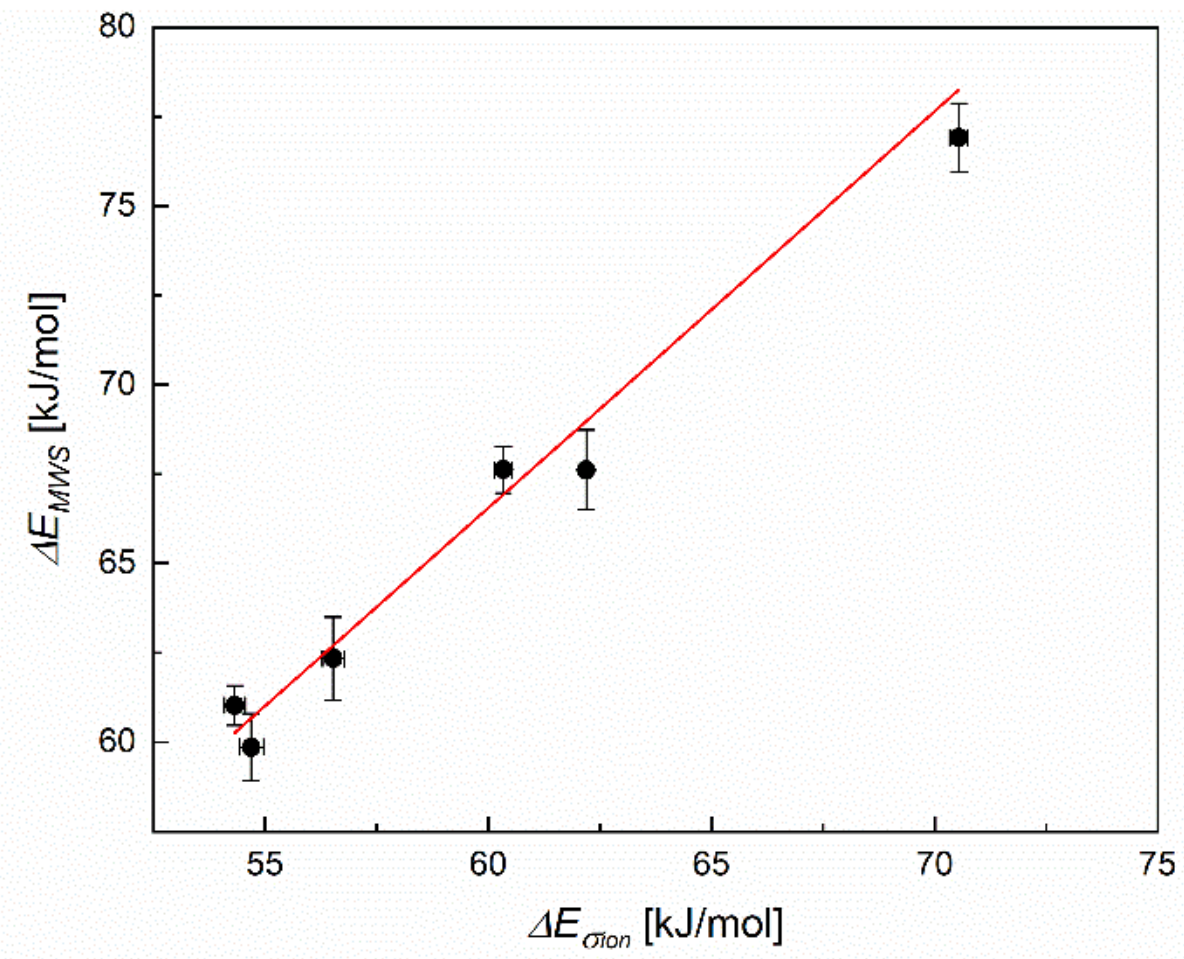

Figure S17. Correlation between activation energies of Maxwell-Wagner-Sillars process and ionic conductivity in the $SmC_A$* phase of MIXmHH6.

Table S3. Fragility index $m_f$ and glass transition temperature $T_g$ obtained for MIXmHH6 based on VFT and MYEGA formulas fitted to the α-relaxation time. Values for m = 2 were obtained based on the Arrhenius fit ($T_V$ or $H$ = 0). Results for high and low temperatures for m = 3, 4, 5 are denoted by h$T$ and l$T$, respectively.

| m | 2 | 3 | 4 | 5 | 6 | 7 |
|---|---|---|---|---|---|---|
| $m_f$, VFT | 23.7(3) | 247(48), h$T$<br>39.3(6), l$T$ | 130(22), h$T$<br>46.5(9), l$T$ | 98(5), h$T$<br>93(4), l$T$ | 69(1) | 90.9(6) |
| $T_g$, VFT | 184(1) | 251(3), h$T$<br>193.1(3), l$T$ | 237(3), h$T$<br>201.5(3), l$T$ | 227.3(8), h$T$<br>219.9(2), l$T$ | 216.7(1) | 221.8(1) |
| $m_f$, MYEGA | 23.7(3) | 118(6), h$T$<br>36.8(4), l$T$ | 86(5), h$T$<br>42.7(5), l$T$ | 75(2), h$T$<br>75.5(5), l$T$ | 61.1(3) | 76.8(2) |
| $T_g$, MYEGA | 184(1) | 242(1), h$T$<br>192.3(2), l$T$ | 232(1), h$T$<br>200.6(2), l$T$ | 224.1(6), h$T$<br>218(2), l$T$ | 215.6(4) | 220.5(2) |

Table S4. Fitting parameters of the $\exp(((T_0 - T)/A)^c)$ function to the ratio of relaxation times of $P_{hex}$ and $P_H$ processes in the $SmX_A$* phase of MIXmHH6. The onset and peak temperatures determined by DSC for the $SmC_A$* → $SmX_A$* transition on cooling are given for comparison.

| m | 2 | 3 | 4 | 5 | 6 | 7 |
|---|---|---|---|---|---|---|
| $c$ | 0.32(5) | 0.13(1) | 0.145(5) | 0.127(4) | 0.182(6) | 0.127(6) |
| $A$ [K] | 5.4(4.5)·$10^{-2}$ | 8.1(7.9)·$10^{-5}$ | 2.3(9)·$10^{-4}$ | 8.5(3.1)·$10^{-5}$ | 2.3(7)·$10^{-4}$ | 7.0(4.1)·$10^{-5}$ |
| $T_0$ [K] | 305.1(5) | 311.5(3) | 309.2(2) | 307.1(1) | 303.3(2) | 298.3(4) |
| $T_o^{DSC}$ [K] | 306.1(4) | 311.6(4) | 310.7(3) | 307.6(2) | 305.4(3) | 301.6(4) |
| $T_p^{DSC}$ [K] | 302.4(2) | 307.7(3) | 306.6(3) | 303.4(3) | 300.5(1) | 295.7(2) |